\documentclass[12pt,preprint]{aastex61}
\usepackage{CJK}
\usepackage{longtable}
\shortauthors{} 
\received{}
\usepackage{graphicx}
\graphicspath{{}}

\begin{document}

\title{On the Origin of the 3.3 Micron Unidentified Infrared Emission Feature}

\CJKtilde
%\begin{CJK*}{UTF8}{gbsn}

\begin{CJK*}{Bg5}{bsmi}
\email{sunkwok@hku.hk}

\author{Seyedabdolreza Sadjadi}
\affiliation{Laboratory for Space Research, Faculty of Science, The University of Hong Kong, Pokfulam Road, Hong Kong, China}

\author{Yong Zhang}
\affiliation{Laboratory for Space Research, Faculty of Science, The University of Hong Kong, Pokfulam Road, Hong Kong, China}
\affiliation{Department of Physics, The University of Hong Kong, Hong Kong, China}
\affiliation{School of Physics and Astronomy, Sun Yat-sen University, Zhuhai 519082, China} %remove this if not appropriate.

\author{Sun Kwok}
\affiliation{Laboratory for Space Research, Faculty of Science, The University of Hong Kong, Pokfulam Road, Hong Kong, China}
\affiliation{Department of Earth Sciences, The University of Hong Kong, Hong Kong, China}
\affiliation{Visiting Professor, Department of Physics and Astronomy, University of British Columbia, Vancouver, B.C., Canada}

\begin{abstract}
The 3.3 $\mu$m unidentified infrared emission feature is commonly attributed to C--H stretching band of aromatic molecules.  Astronomical observations have shown that this feature is composed of two separate bands at 3.28 and 3.30 $\mu$m and the origin of these two bands is unclear.  In this paper, we perform vibrational analyses based on quantum mechanical calculations of 153 organic molecules, including both pure aromatic molecules and molecules with mixed aromatic/olefinic/aliphatic hydridizations.
We find that many of the C--H stretching vibrational modes in polycyclic aromatic hydrocarbon (PAH) molecules are coupled.
 Even considering the un-coupled modes only, the correlation between the band intensity ratios and the structure of the PAH molecule is not observed and the 3.28 and 3.30 $\mu$m features cannot be directly interpreted in the PAH model.  %an exact identification of the astronomical feature is not trivial.  
Based on these results, the possible aromatic, olefinic and aliphatic origins of the 3.3 $\mu$m feature are discussed.   We suggest that the 3.28 $\mu$m feature is assigned to aromatic C--H stretch whereas the 3.30 $\mu$m feature is olefinic.  From the ratio of these two features, the relative olefinic to aromatic content of the carrier can be determined.          
\end{abstract}


\section{Introduction}

The 3.3 $\mu$m emission feature was first discovered in the planetary nebula NGC 7027 \citep{merrill1975}.  Extension of spectral observations to longer wavelengths revealed that the 3.3 $\mu$m features is a member of a family of strong bands at 3.3, 6.2, 7.7, 8.6 and 11.3 $\mu$m \citep{RSW77}, which are now collectively called  unidentified infrared emission (UIE) bands.  Since these features are too broad to be atomic lines and without any substructures to qualify as molecular bands, they were believed to arise from mineral solids \citep{RSM77}.  This seemed a reasonable interpretation at the time because silicate minerals have already been found to be common among evolved stars \citep{woolf1969}.  

Interestingly, the 3.3 $\mu$m UIE feature was almost immediately proposed as originating from the C--H stretching mode of organic compounds \citep{knacke, duley1979}.  However, the organic interpretation was largely ignored as organics were considered to be an unlikely component of the interstellar medium at that time.  Only after the discovery of a large number of simple gas-phase organic molecules in space through their rotational transitions in the millimeter-wave band was organic compounds being taken seriously by the astronomical community.  For the last 30 years, the UIE bands have been commonly attributed to radiatively excited vibrational modes of polycyclic aromatic hydrocarbon (PAH) molecules \citep{allamandola1989, puget1989}.  

In spite of the popularity of the PAH model, there are major problems associated with this hypothesis \citep{kwok2011, kwok2013}. It is well known among chemists that the vibrational modes of PAH molecules spread over wide range of wavelengths and no specific PAH molecules match the observed wavelengths and band intensities of the astronomical UIE bands \citep{cook1996, cook1998, Wagner00}.  The detection of  3.4 $\mu$m \citep{merrill1975} and 6.9 $\mu$m \citep{puetter1979} bands in many UIE sources also complicates matter.  These two features can be traced to C--H stretch and C--H in-plane-band of aliphatic compounds \citep{duley1981}.  Although the 3.3 $\mu$m band is generally much stronger than the 3.4 $\mu$m band, in some sources (e.g., proto-planetary nebulae) the two features can be comparable in strength \citep{geballe1992, hrivnak07, Goto07}.

The laboratory measurements of the 3.3 and 3.4 $\mu$m features dated back to 1905, when  organic molecules containing normal saturated hydrocarbons and benzene were studied with the technique of infrared absorption spectroscopy  \citep{Coblentz1905}. 
The first identification of these bands as due to vibrations of molecules %such as methyl (CH$_{3}$) and methylene (CH$_{2}$) groups 
were provided by mechanical models with empirical force constants  \citep{Martin1937, Martin1938, Martin1939, Martin1940}.  These interpretations were later confirmed by quantum mechanical {\it ab initio} electronic structure calculations  \citep{Hehre1986}.

Higher spectral resolution observations of the astronomical 3.3 $\mu$m feature shows that it is in fact composed of two components: a feature peaking at 3.28 $\mu$m and another at 3.30 $\mu$m \citep{tokunaga1991, Song03, Hammonds15}.  The observation that the two components are emitted in different spatial locations suggest that these two components originate from different (although possibly related) chemical species \citep{Candian12}.

Two explanations have been proposed for the vibrational nature of the 3.3 $\mu$m feature.
One is based on the PAH model where the 3.28 and the 3.30 $\mu$m components arise from `bay' and `non-bay' hydrogen sites of the PAH units  \citep[see Figure 3 of][]{Candian12}.
This interpretation is  mainly based on the detection of two bands within the range of 3.33$-$3.21 $\mu$m in the experimental gas-phase infrared spectra of some small PAH molecules with bay-type hydrogens in their molecular structures \citep[see Figure 6 of][]{Candian12}. 
Alternatively,  \citet{Chiar13} assign an aliphatic origin to the 3.28 $\mu$m component and interpret it as the stretching mode of olefinic C$-$H bonds in amorphous hydrocarbons.

In the PAH hypothesis, the 3.4 $\mu$m feature is the result of superhydrogenation of PAH molecules \citep{sch93}. 
It has also been interpreted as arising  from hot bands from anharmonic aromatic C--H stretch, which shifts the 3.3 $\mu$m feature to a longer wavelength  \citep{barker}.   However, theoretical calculations including anharmonicity show a simultaneous increase of the width of the 3.3\,$\mu$m feature, which was not observed \citep{van04}. Furthermore, the expected strong overtone bands at the 1.6--1.8\,$\mu$m region is not detected, making the anharmonicity explanation unlikely \citep[][and references therein]{goto03}. 
%Nevertheless, anharmonicity effect may contribute to spectral activities  in the 3.3\,$\mu$m region \citep{mal15,mal16}.

Most likely, the astronomically observed 3.4 $\mu$m feature is aliphatic in nature and arises from C$-$H stretching modes of methyl and methylene functional groups  \citep{Sandford1991,Goto07, SKZ2015-1}. Distinct components of the 3.4 $\mu$m have been observed at 3.40, 3.46, 3.52, and 3.56 $\mu$m \citep{Jourdain86, Jourdain90, hrivnak07}.
%However, observationally this feature can extend as far as 3.5 $\mu$m  \citep{Sellegren2001,Van02,Habart2004}, 

Another class of molecules which C--H stretching modes can contribute to the observed 3.3 $\mu$m bands are diamondoids.  For example, 
adamantane (C$_{10}$H$_{16}$), with a cage-like, full $sp^{3}$ structure, shows vibrational bands in the 3 $\mu$m region in both gas and solid phases \citep{Oomens2006,Pirali2007}. 
%These molecules while composed of methylene groups i.e the same building blocks of linear normal alkanes, they are accompanied by tertiary $sp^{3}$ C$-$H groups with the slightly longer C$-$H bond lengths (theoretical value:0.00336 \AA\ \citep{Oomens2006}). Such tertiary groups are frequently found in hydrocarbons with branch-like structures. It is still an open question Which types of these C$-$H bonds are responsible for the appearance of 3.5$\mu$m feature in diamondoids. 
The C--H stretch of hydrogenated diamonds at 3.43 and 3.53 $\mu$m have been detected \citep{guillois1999}.  After the discovery of fullerenes (C$_{60}$) in the interstellar medium, hydrogenated fullerenes (fulleranes) have also been detected through their C--H stretching modes \citep{zhang2013}.

At an even higher complexity level, many amorphous hydrocarbons show strong spectral features around the 3.3 $\mu$m region.  The best known examples are hydrogenated amorphous carbon (HAC) \citep{scott1996} and quenched carbonaceous composites (QCC) \citep{sakata1987}.  
The mixed hybridization nature of these materials produces a richer spectrum in the 3.3 to 3.5 $\mu$m region \citep{Wada2006}.
%also show both 3.4 and 3.5$\mu$m \citep{Wada2006} attributed to the formation of different types of C$-$H bonds (due to the fully or partially hydrogenation of the carbon atoms) at the edge of the solid \citep{Wada2006}. 

When heteroatoms are added to the hydrogen-carbon mix, other spectral features can appear.
%In some circumstances, the electronic effects of neighboring groups can alter the frequency of stretching of C$-$H bonds far beyond the standard values. 
A good example is the $sp^{2}$ C$-$H bond in carbonyl group of aldehyde family %(with the carbonyl group as the electron withdrawing group) 
with its stretching mode lies within the range of 3.45$-$3.77$\mu$m  \citep{Socrates2001}.  
%This gives the possibility to detect biologically important molecules spread in the interstellar media.
Replacing the carbon atom with the nitrogen in a neutral PAH molecule can cause blue shifts of the 3.3 $\mu$m  band  \citep{Mattioda2017}. 

%Thus the presence of the heteroatoms in the molecular formula of the hydocarbons can be well detected in this range of IR wavelength.

Carbon is a soft atom with the ability to make the variety of classical and non-classical molecular structures and C$-$H bonds  \citep{Minyaev2008}. The stretching vibration of C$-$H bond in different hybridization and singlet or triplet electronic states spans the wavelength range of 2.8 to 3.7 $\mu$m.
It is clear from the above discussions that the origin of the 3.3 $\mu$m UIE band is far from settled.  
In this paper, we explore the possible origins of the 3.3 $\mu$m feature through quantum chemical vibrational analysis of 153 molecules ranging in structure from pure PAH molecules to molecules with aromatic cores and aliphatic side groups and pure aliphatic chains.

\section{Method}

Our calculations are based on density functional theory   B3LYP  \citep{becke1993a,hertwig1997} functions and BHandHLYP hybrid functionals \citep{becke1993b}  in combination with polarization consistent basis set PC1   \citep{jensen2001,jensen2002}.  
We obtain the local minimum geometries and the harmonic frequencies of fundamental vibrations of four groups of molecules: PAH (Figure \ref{pah}), ally-PAH (Figure \ref{ally}), alkyl-PAH (Figure \ref{alkyl}) and normal and branched alkane organic compounds (Figure \ref{alkane}).  The names and molecular formulas of these molecules are listed in Table \ref{bay-PAH}--\ref{aliphatics}. 
The double scaling factors scheme of  \citep{laury2012} are applied to the DFT harmonic vibrational frequencies. In this scheme the harmonic frequencies $>1000$ cm$^{-1}$ and $<1000$ cm$^{-1}$ are scaled by 0.9311 and 0.9352 for BHandHLYP hybrid functionals  and 0.9654 and 0.9808 for B3LYP, respectively.

The calculations were performed  using the Gaussian 09, Revision C.01 software package  \citep{fri09} running on the  HKU grid-point supercomputer facility.  The B3LYP calculations were done using PQS \footnote{PQS version 4.0,  Parallel Quantum Solutions,   2013 Green Acres Road,  Fayetteville,  Arkansas  72703   URL: http://www.pqs-chem.com  Email:sales@pqs-chem.com: Parallel Quantum Solutions.} running on QS128-2300C-OA16 QuantumCubeTM machine.
Under the default criteria of both software, all the optimized geometries were characterized as local minima, established by the positive values of all harmonic frequencies and their associated eigenvalues of the second derivative matrix.

Visualization and manipulation of the results of vibrational normal mode analysis were performed by utilizing the Chemcraft \footnote{http://www.chemcraftprog.com} suit program. 
Quantitative vibrational analysis on atomic displacement vectors were performed by Vibanalysis code V2.0, as described in \citep{SKZ2015-1}. 
In order to simulate astronomical spectroscopic observations, 
a Drude broadening profile of $T$=500 K and FWHM=0.03 is applied to the experimental resolved or theoretically calculated vibrational transitions  \citep{SKZ2015-1}.  The value of the FWHM is chosen to match the FWHM value of the astronomical UIE features at 3 to 4 $\mu$m region of IR spectra \citep{hsia2017}.

%\subsection{Accuracy of the Calculations}

The accuracy of the theoretical calculations can be tested by comparison with gas-phase laboratory infrared spectra  \citep{Zvereva2011}. 
Previously, we have estimated errors of 0.12--0.13 $\mu$m for our density functional theory calculations \citep{SKZ2015-1}. 
We have compared the  experimental spectra  (2 $\mu$m $\leq$ $\lambda$ $<4$ $\mu$m) of  28 PAH molecules in the version 2 of the NASA-Ames PAH data base \citep{boersma} with a test set of 60 neutral PAH molecules \citep{SKZ2015-2} and the results are shown in Table \ref{DFT-PAH}. 
We find average errors of 0.03316 $\mu$m and 53 kcal/mol in wavelength and absolute intensity, respectively.  
It should be noted that although DFT calculations show  large error in predicting the absolute intensity values, they can predict good relative intensities \citep{Zvereva2011, SKZ2015-1}.  

\section{The 3.3 micron Feature}
\subsection{Vibrational analysis of `bay' and `non-bay' aromatic C$-$H stretching modes}

\begin{figure}
	\begin{center}
		\resizebox{175mm}{!}{\includegraphics{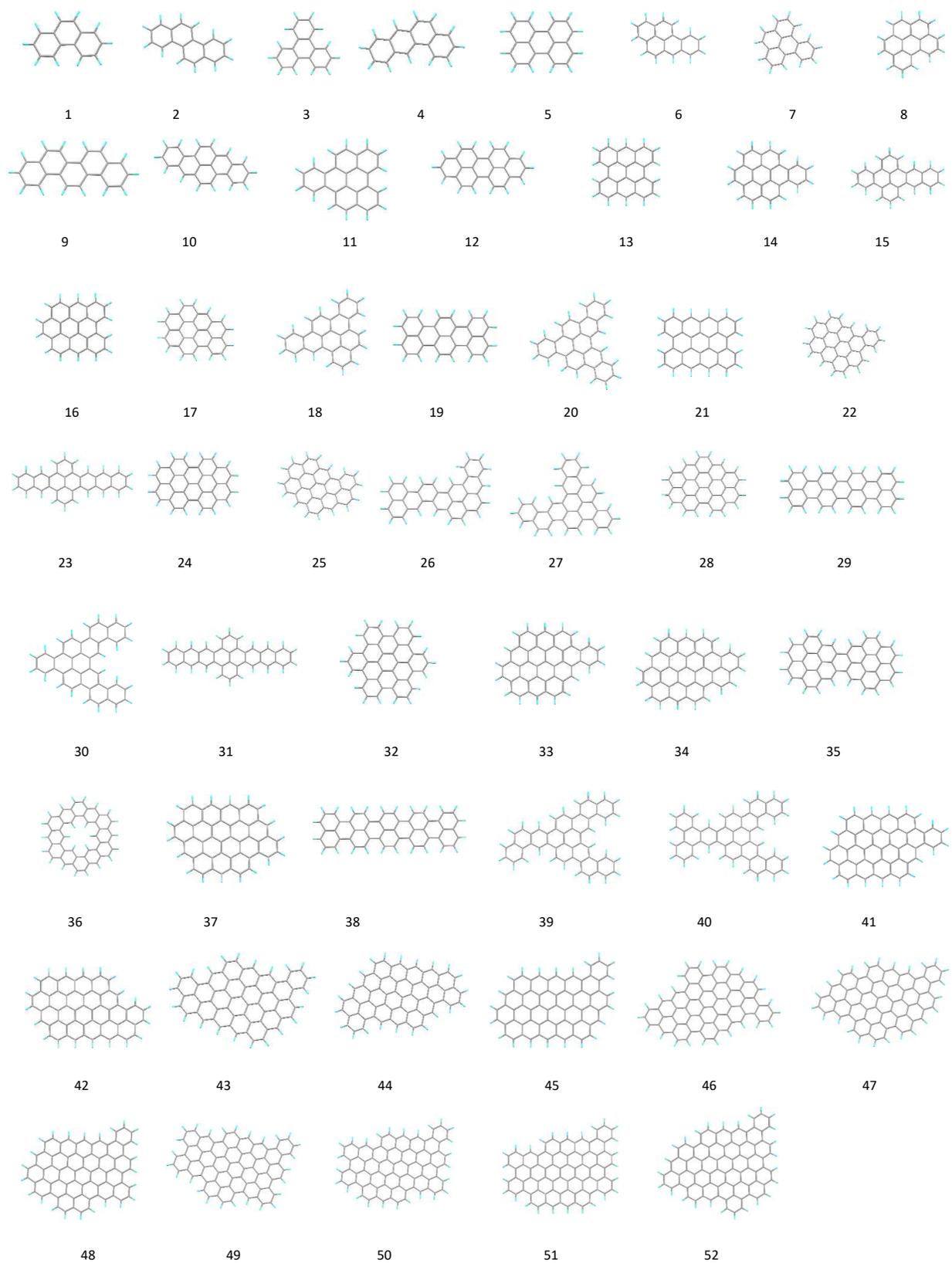}}
	\end{center}
	\caption{\label{pah} Local minimum geometries of 52 planar neutral PAH molecules with `bay' C--H sites.}
	
\end{figure}

In order to quantitatively evaluate the `bay' and `non-bay' origins for the 3.28 and 3.30 $\mu$m features as proposed by \citet{Candian12}, we have computed the vibrational modes of a group of 52 neutral PAH molecules of different sizes and shapes with peripheral hydrogen atoms in `bay' configurations  (Figure \ref{pah}, Table \ref{bay-PAH}). The relative numbers of `bay' C$-$H bonds to the total aromatic C$-$H bonds in each of these molecules are listed under column 4 of Table \ref{bay-PAH}. The number of the normal modes for each molecule listed in Table \ref{bay-PAH} is equal to the number of aromatic C$-$H bonds or simply the hydrogen atoms in the molecular formula.  
In total, there are 986 normal modes for the 52 molecules, and these modes (including both Raman and IR active modes)
are analyzed based on the  vibrational displacement vectors in the calculations \citep{SKZ2015-1} in which we use criterion that atoms with total displacement of less than 0.01\AA\ are set as stationary atoms with zero contribution to the normal mode.
%The dependence of the peak wavelengths of the vibrational modes on the fraction of bay C--H bonds are plotted in Figure \ref{bay}. 

%The wavelengths of the vibrational transitions with sole contribution of bay C$-$H bonds covers a broad range from 3.16 to 3.25 $\mu$m (right side of Figure \ref{bay}), whereas the solely non-bay vibrations cover a range from 3.27 $\mu$m from 3.28 $\mu$m (between the two horizontal lines Figure \ref{bay}). 

Figure \ref{bay} shows the degree of contribution from `non-bay' (top panel) and `bay' (bottom panel) C$-$H bonds to vibrational modes at each frequency in the 3 $\mu$m region.  These results show a considerable coupling of both `bay' and `non-bay' aromatic C$-$H stretching bond vibrations. Approximately half of the transitions are pure `bay' (111) and `non-bay' (372) and the remaining (503) are coupled transitions.
%We can see that most modes have a mixed contribution, or the bay and non-bay C$-$H bonds are coupled.  
If we ignore these couplings, the vibrations with sole contribution from `bay' C$-$H bonds appear at shorter wavelengths than the `non-bay' vibrations. 
The `bay' C--H bond lengths are generally slightly shorter than `non-bay' bond lengths, thus their vibrations are expected to appear at slightly shorter wavelengths.
Although this is qualitatively consistent with the suggestion of \citet{Candian12}, the considerable coupling suggests  that we would be unable to directly link the `bay' and `non-bay' aromatic C$-$H vibrations to the 3.28 and 3.30 $\mu$m astronomical components. 

%\subsection{Relative Intensity of Bay and Non-bay aromatic C$-$H stretching}    

From astronomical observations, the flux ratio of the 3.28/3.30 $\mu$m varies from 0.36 to 4.97 \citep{hsia2017}. Can this range of flux ratios be explained by `bay' and `non-bay' C$-$H bonds?  
%Both `bay' and `non-bay' C$-$H bonds are aromatic with $sp^{2}$ carbon atoms but the bond lengths for the `bay' C$-$H bonds are slightly shorter than `non-bay' ones. For example, in phenanthrene the length for `bay' bonds is 1.07785 \AA\ versus 1.07998 \AA\ for `non-bay' bonds.
%If we assume that the electric moments and polarizabilities are the same and ignore all complex coupling effects in vibrations, then the flux ratio of 3.28/3.30 $\mu$m should correlate linearly (positive slope) with the ratio of number of `bay' C$-$H bonds to `non-bay' bonds as listed in column 4 of Table \ref{bay-PAH}.  Out of the 111 pure `bay'  normal modes and 372 `non-bay' normal modes transitions shown in Figure \ref{bay},  the average values of intensities per C$-$H bonds are 2.17367 and 1.55513 km/mol for `bay' and `non-bay' bonds respectively.
The relative intensities of the vibrational modes are convolved with a Drude profile at $T$=500 K and the band profiles of the `bay' and `non-bay' transitions are shown  in Figure \ref{drude1}. 
These results suggest a theoretical band flux ratio (F$_{bay}$/F$_{non-bay}$) of 1.17.

The separation between the peak of the `bay'  and the `non-bay' profiles is  0.054 $\mu$m, which is larger than the astronomical observed wavelength difference  between 3.28 and 3.30 components. 
%This difference could be the result of coupled modes as seen in Figure \ref{bay}.  
For comparison with the theoretical calculations, we have also produced simulated astronomical spectra from laboratory data of PAH molecules with `bay' and `non-bay' C$-$H bonds in their structures (Figure \ref{lab}).    There is a slight difference between the peak wavelengths of the 3 $\mu$m C--H stretch of the two groups, but the difference is not large enough to separate the molecules based on the spectra alone.  This suggests that there is a high degree of coupling between the `bay' and `non-bay' vibrations in actual PAH molecules.

To further examine the difference in peak wavelengths and strength ratios between the `bay' and `non-bay' C--H bonds, we separate the PAH molecules in Figure \ref{pah} into three groups according to their bay\% values in Table \ref{bay-PAH}.  The three groups correspond to bay percentages of $<$50\%, =50\% and $>$50\%, i.e from low fractions of bay C$-$H in molecular structure to high. 
%This classification is labeled as scheme 1. 
In each of these groups the transitions with pure `bay' and `non-bay' contributions are separated and their corresponding infrared bands are simulated by Drude model at $T$=500 K. These profiles are plotted in Figure \ref{scheme1}.  This figure shows that in all classes the band associated with the  stretching modes of `bay' C$-$H bonds appears at shorter wavelength than `non-bays'. However the wavelength difference between two bands changes in a complex way. 
It increases from 0.047 $\mu$m in $<$50\% bay class to 0.06 $\mu$m in $>$50\% bay class. The changes in the bands relative flux values are also not simple. The almost similar values of relative flux values of bay/non-bay bands in PAH molecules with $>$50\% and =50\% bay make these two classes indistinguishable.  We do not observe any linear correlation in bands flux ratio values with an increase in percentage number of `bay' C$-$H bonds.  
The variation of the values of band flux ratio in three different classes of molecules lies within a much narrower interval (0.948–-2.809) than the observed range of 0.36--4.36.  The only case that the band flux ratio is smaller than 1.00 is the class with $<$50\% bay characteristics.   It appears that `bay' and `non-bay' PAH molecules will not be able to reproduce the wide range of ratios as observed.

In order to test whether the observed 3.25--3.3\,$\mu$m flux ratios can reflect the number ratio of `bay' and `non-bay' PAH sites,  we have listed in columns 5 and 6 of Table \ref{bay-PAH} the  band strengths per unit `bay' and `non-bay' C--H bonds ($A_{\rm bay}$ and $A_{\rm nonbay}$).  These values are calculated from the sum of the intensities ($I_{n}$) of all pure `bay' or `non-bay' transitions in a PAH molecule divided by the total number ($N_{bay}$) of `bay' or `non-bay' C--H bonds within the structure.  
One may expect to infer from the observed intensities ($I_{\rm bay}$ and $I_{\rm nonbay}$) the number ratios between `bay' and `non-bay' C--H bonds %($R_{\rm b/n}$) 
from $(I_{\rm bay}/I_{\rm nonbay})\times(A_{\rm nonbay}/A_{\rm bay})$.  The last column of Table \ref{bay-PAH} lists the calculated
$A_{\rm nonbay}/A_{\rm bay}$ ratios, which turn out to be significantly different from one molecule to another.
The variations  in the values of $A_{bay}$, $A_{non-bay}$ and $A_{bay}$/$A_{non-bay}$ among different PAH molecules are due to the fact that just a certain numbers of C--H bonds (not all of them simultaneously) participate in the vibrations of each normal mode.  As a result, the band strengths cannot be directly linked to the simple number of bay or non-bay sites.
%Thus this way of linking the band intensities to the structure can not correctly reflects the observational bands flux ratio.
Combined with the coupling effect, the profile of the 3.25--3.3\,$\mu$m feature does not simply reflect the structures of PAH molecules.  Because of these complications, it would be simplistic and  misleading if one calculates the the number ratios between `bay' and `non-bay' through such an approach.

%Thus ultimately we anticipate that the observed variation in the band flux ratio (3.28/3.3 µm) in the sample of astronomical resources is likely due to the existence of other functional groups in the structure of the unknown interstellar organic molecules
%The last paragraph of Section 3.1: The last sentence is incorrect. The experimental data are those of PAHs with known structures (not astronomical species). The different wavelength separations can support the coupling scenario, but cannot give any implication on "another functional group".
%It is quite possible that another functional group is needed to explain the observed subfeatures.
     
\subsection{Vibrational analysis of olefinic $sp^{2}$ C$-$H stretching mode}

In this  section we examine the alternative assumption on the olefinic origin of the 3.28 $\mu$m component \citep{Chiar13}.
A group of 62 allyl-aromatic molecules are created by adding one  to three H$_2$C=CH$-$CH$_2-$ groups to a PAH molecular core (Figure \ref{ally}).    Each molecule in this set  contains three types of C$-$H bonds: aromatic $sp^{2}$,  aliphatic $sp^{2}$ (olefinic), and aliphatic $sp^{3}$ bonds. 

A total number of 1248 vibrational normal modes including infrared and Raman active modes are analyzed. The contributions of each of these three types of C$-$H bonds vibrations are plotted against their corresponding wavelengths in Figure \ref{contributions}.   
We can see that the aliphatic $sp^{3}$ (top panel) and aromatic $sp^{2}$ (bottom panel) C$-$H stretching vibrations form  well-defined bands.
However, the  olefinic C$-$H stretching mode (middle panel) forms three separate groups, centering around 3.22, 3.30 and 3.35 $\mu$m. 
The 3.22 and 3.30 $\mu$m bands of olefinic C$-$H vibrations contain pure olefinic C$-$H bonds vibrations (100\% contributions) with zero couplings to either aromatic or  aliphatic $sp^{3}$ C$-$H stretching modes.  We note that parts of these two bands overlap in wavelength with aromatic transitions (bottom panel). 

The 3.35 $\mu$m band consists of olefinic stretching modes coupled with aliphatic $sp^{3}$ vibrations. The average contributions of olefinic stretching modes associated to this band is 30\%.

%\subsection{Relative contributions from aromatic, olefinic and aliphatic C$-$H stretching to the 3 micron band}

The relative contributions from the three different kinds of C--H stretch to an astronomical 3 $\mu$m feature are shown in  Figure \ref{pure}.   We can see that the olefinic C--H stretch has a blue shoulder that overlaps with the aromatic C--H stretch.  Olefinic C$-$H stretching modes therefore can contribute to the astronomically observed 3.3 $\mu$m feature.  
%{\bf Section 3.2, last paragraph, line 2: We can see…has a strong peak at 3.3 um and a blue shoulder...; We may mention that that aliphatic/aromatic strength ratio (1.77) is in good agreement with that obtained by Yang et al. (2013; 1.76).}
Among the three stretching modes, the aliphatic C--H stretch is about twice as strong as 
the aromatic and olefinic C$-$H modes in these molecules (Figure \ref{pure}). 
%Our calculated (F$_{ali}$/F$_{ar}$ =1.77) is in an excellent agreement with the value of (A$_{ali}$/A$_{ar}$ =1.76) reported by  \citet{Yang2013}. Although the types of molecules, the numbers of aliphatic, aromatic C$-$H bonds and also the basis sets applied are different in our work and  \citet{Yang2013} the excellent agreement between two reports confirm the intrinsic IR activity of these two type of C$-$H bonds.

\subsection{Mixed aromatic-aliphatic structures}

Next we study a group of 21 alkyl-PAH molecules with a larger aliphatic component (Figure \ref{alkyl} and Table \ref{alkyl-PAH}). A single aliphatic branch is added to an aromatic core of multiple rings.  Each of these molecules have equal numbers of aliphatic and aromatic C$-$H bonds in their structures. 
If there is zero degree of coupling between the vibrations of two types of C--H bonds (Figure \ref{contributions}), the flux ratio of simulated bands in all these structures should follow the intrinsic flux ratio of aliphatic/aromatic calculated in Figure \ref{pure}.

The simulated infrared spectra of this group of alkyl-PAH molecules are shown in Figure \ref{drude_alkyl}.  Since the aliphatic C--H stretch has larger intrinsic strength than the aromatic C--H stretch, the 3.4 $\mu$m band is more prominent than the 3.3 $\mu$m band.  In order to produce a relatively stronger 3.3 $\mu$m band, the fraction of aromatic/olefinic components must be larger in these molecules.  

%We also looked at the available experimental data for such types of molecules (Figure \ref{ethyl}). 

%From our vibrational analysis results on zero coupling between aliphatic and aromatic C$-$H vibrations, it is concluded that the band pairs (3.253 ,3.296 $\mu$m) and (3.367, 3.455 $\mu$m) contain pure aromatic and aliphatic $sp^{3}$ C$-$H stretching transitions (Figure \ref{ethyl}). The total flux of the aliphatic bands for ethylbenzene (Figure \ref{ethyl}) is calculated to be 1.7 times greater than those of aromatic bands. Once again our theoretical estimation of (F$_{ali}$/F$_{ar}$ =1.77) is validated.

%These data are confirming that our theoretical estimation of relative band fluxes associated to three types of C$-$H bonds in the condition of zero coupling between them are reliable.    

\section{The 3.4 micron feature}
%\subsection{Vibrational Analysis on Methyl and Methylene C$-$H Stretching modes}

The astronomical 3.4 $\mu$m feature is commonly attributed to methyl ($-$CH$_{3}$) and methylene ($-$CH$_{2}-$) C--H stretching modes  in the 3.37$-$3.51 $\mu$m region. This region is subdivided into four segments : anti-symmetric C$-$H stretching vibration in methyl groups (3.37$-$3.39 $\mu$m), anti-symmetric C$-$H stretching vibration in methylene groups (3.41$-$3.43 $\mu$m), symmetric C$-$H stretching in methyl group (3.47$-$3.50 $\mu$m) and symmetric C$-$H stretching in methylene group (3.49$-$3.51 $\mu$m)  \citep{Colthup1990}.

In order to study these vibrations, we have created  20 hydrocarbons (10 linear and 10 branched) with up to 50 carbon atoms (Figure \ref{alkane} and Table \ref{aliphatics}). All these molecules are composed of only methyl  and methylene  groups without any tertiary $-$CH groups as in 2-methylpropane.
The results of our quantitative analysis on the 594 vibrational modes in these hydrocarbons are presented in Figure \ref{symm}.
The vibrations of methyl and methylene C$-$H  bonds are clearly separated  based on their contributions in each normal modes and the type of vibrations (symmetric or anti-symmetric). 

It is observed that the couplings between all four types of  vibrational motions are significantly large (Figure \ref{symm}). The most significant coupling occurs at the wavelength range of 3.4$-$3.44 $\mu$m where all types of methyl and methylene vibrations have non-negligible contributions in normal mode vibrations.
Only within the narrow range of wavelength between 3.45$-$3.47 $\mu$m that the pure un-coupled vibrations of symmetric C$-$H stretching of methyelene groups are observed (second panel from top in Figure \ref{symm}). This group of transitions defines the long wavelength boundary of 3.4 $\mu$m features.  
%Any features detected beyond 3.47 $\mu$m (e.g., the 3.56 $\mu$m feature) cannot be due to methyl/methylene C--H stretching motions.

Figure \ref{uncoupled} shows the wavelength peaks of the vibrational modes if the un-coupled transitions (100\% contribution) are selected for each class of methyl and methylene vibrations.  The pattern  is in perfect agreement with that of \citet{Martin1938} where the blue and red shifted boundaries are composed of anti-symmetric C$-$H stretching mode in methyl and symmetric C$-$H stretching mode in methylene groups respectively.
Although the four peaks of the symmetric and anti-symmetric stretches of the methyl and methylene groups are clearly separated, they cannot be mapped one-to-one to the astronomically observed 3.4 $\mu$m bands at 3.40, 3.46, 3.52, and 3.56 $\mu$m \citep{hrivnak07}.  Since the C--H stretching frequencies (symmetric or anti-symmetric) of methyl and methylene groups do not go beyond 3.5 $\mu$m, the observed 3.56 $\mu$m (or even the 3.52 $\mu$m) feature cannot be due to these vibrational modes.

The methyl group vibrations are often characterized by the vibrations of all of its three C$-$H bonds, which are called  ``trio'' vibrations. In our calculations, we can also observe and separate other subclass vibrations for  methyl groups involving only two (``duot'') or one (``solo'') motions of methyl C$-$H bonds (Figure \ref{ch3}). We can see that the major vibrational characteristics of the methyl's C$-$H vibrational modes are trio and duot types of stretching. 
When the methyl group rotates fast around single C$-$C bond at high temperatures 
\citep[above 2.9 kcal.mol$^{-1}$,][]{Payne1977} the  C$-$H stretch of the group is mostly trio. 

 \section{Discussion}

In a theoretical study  of several pyrene-like and perylene-like PAH molecules with ``armchair'' edges, \citet{can14} found two strong peaks at 3.23 and 3.26\,$\mu$m and attributed them to symmetric stretching of C--H bonds involving duo hydrogens and anti-symmetric
stretching  of C--H bonds involving duo and trio hydrogens. 
Our calculations for a large set of molecules show that pure `bay' and `non-bay' C--H stretching modes results in two peaks at 3.20 and 3.26\,$\mu$m, respectively (Figure \ref{drude1}). This is qualitatively consistent with the finding of \citet{can14} since the PAHs with more C--H bonds involving duo hydrogens have more bay sites (see their Figure~1).
\citet{can14} also proposed that the armchair PAHs are responsible for the 12.7\,$\mu$m band, and suggested that there is a correlation between the spatial distributions of the 12.7\,$\mu$m band and the two components of the 3.3\,$\mu$m bands. However, our results indicate that strong coupling of `bay' and `non-bay' vibrations may blur such a correlation. 

%In principle, less compact PAHs will introduce more bay sites and thus may lead to a rising of the blue part of the 3.3 µm feature. From the observations of the Red Rectangle, Candian et al. (2012) found that the 3.3 µm feature exhibits rising blue part with increasing distance from the central star, and attributed this to the growth of PAHs by adding C$_4$H$_2$ entities. However, a quantitative interpretation is no easy due to......

In principle, less compact PAHs will introduce more bay sites and thus may lead to a rising of the blue part of the 3.3\,$\mu$m feature.
From the observations of the Red Rectangle,  \citet{Candian12} found that the 3.3\,$\mu$m feature exhibits rising blue part with increasing distance from the central star, and attributed this to the growth of PAHs by adding C$_4$H$_2$ entities.   However, this correlation cannot be easily interpreted due to the  non-linear relation between 3.2 $\mu$m/3.26 $\mu$m intensity ratios and the bay/non-bay number ratios.

Our results show that the theoretical peaks of both `bay' and `non-bay' C--H stretching modes are located in shorter wavelengths than the observed 3.3\,$\mu$m band. In contrast, the olefinic C--H vibration shows a strong peak at 3.3\,$\mu$m, while the aromatic C--H vibration manifests itself as a blue shoulder (Figures~10 and 11). It seems to be more reasonable to hypothesize that in astronomical spectra the red part of this feature originate from olefinic C--H bond while the blue wing mainly from aromatic C--H bond (with minor contribution from olefinic one).

Because the olefinic and aromatic vibiraitons are decoupled, the intensity ratio of the two components directly reflects their corresponding C--H number ratios. If the aromatic/olefinc hypothesis holds, the rising blue wing with increasing offset from the center of the Red Rectangle nebula
would indicate that the UV photons in the interstellar medium is processing the olefinic component into more stable aromatic ring. The high 3.3 $\mu$m/3.4 $\mu$m intensity ratio detected in many astronomical sources \citep[e.g.][]{Yang2013} may suggest that the UIE carrier has significant olefinic components.

\section{conclusions}

In this paper, we have investigated in detail the  origin of the UIE bands in the 3.3--3.4 $\mu$m region with quantum chemistry calculations. Possible contributions from aromatic $sp^2$, aliphatic $sp^2$ (olefinic), and aliphatic $sp^3$ C--H stretches to the observed bands are studied. Many of these vibrational modes show a significant degree of coupling, so the interpretation of astronomical spectra in this spectral region is not as straight forward as commonly believed.

We confirm that aromatic C$-$H bonds in the form of a `bay'  vibrate at shorter wavelengths than `non-bay' C$-$H.  However the difference is  $\sim$0.05 $\mu$m, which is larger than the astronomical peak separation of 0.02 $\mu$m between the 3.28 and 3.30 $\mu$m features.
Furthermore, the strength ratios  of the `bay' and `non-bay' bands do not correlate in a simple way with the number of bay C$-$H bonds.  In many instances, the `bay' and `non-bay' vibrations are coupled, so the 3.28 and the 3.30 $\mu$m UIE bands cannot be unambiguously identified as `bay' and `non-bay' C--H stretching from PAH molecules.

We propose to assign  the 3.30 $\mu$m feature to olefinic C--H stretch while keeping the assignment of the 3.28 $\mu$m feature as aromatic C--H stretch.  Since the olefinic and aromatic vibrations are intrinsically un-coupled, the ratio of the 3.28 and 3.30 $\mu$m bands could in principle be used to determine the fractions of olefinic groups in the molecule (Figure \ref{pure}).

In the 3.4 $\mu$m region, we show that the un-coupled C$-$H stretching modes of methyl and methylene functional groups can explain the observed features in this region.  However, the reality is more complicated as there exists quite extensive couplings between these modes.  While there is no doubt that the astronomical 3.4 $\mu$m bands are aliphatic C--H stretches, we cannot assign exact identifications to the observed 3.40, 3.46, 3.52, and 3.56 $\mu$m features.  One definite conclusion one can draw is that the 3.56 $\mu$m feature cannot be due to a pure C--H bond and probably involve an element other than C and H (e.g.,  an aldehyde group).

%Finally our vibration analysis on different C$-$H stretching modes of methyl and methylene functional groups revealed a large coupling between the modes associated to them. In the un-coupled situation our results are in complete agreement with all empirical or semi empirical interpretations which are widely accepted and used.         

Although we have extended the vibrational analysis beyond purely aromatic molecules to molecules with aliphatic components, we note that the molecules considered are still quite simple.  We have yet to explore complex organic compounds similar to those of HAC or QCC.  As the size increases and the geometric getting more complex, it is possible that a different qualitative spectral pattern may emerge.  We will continue to investigate molecules with more complex structures as well as those that contain other elements beyond carbon and hydrogen.

{\flushleft \bf Acknowledgements~}
The Laboratory for Space Research was established 
by a special grant from the University Development Fund of the University of Hong Kong. This work is also in part supported by grants 
from the HKRGC (HKU 7027/11P and HKU7062/13P).

\software{Gaussian 09 (C.01; Frisch et al. 2009), PQS (v4.0), Chemcraft, Vibanalysis (V2.0; Sadjadi et al. 2015a)} 

\begin{flushleft}
	
\end{flushleft}

\clearpage

	   \begin{longtable}{ccccccc}
		\caption{Names and molecular formulas for PAH molecules in Figure \ref{pah}} \label{bay-PAH} \\
		
		\hline\noalign{\smallskip}
		No & Name & Formula & Bay\% & $A_{bay}$$^{1, 2, 3}$ & $A_{nonbay}$$^{1, 2, 3}$ & $A_{bay}$/$A_{nonbay}$ \\
		\noalign{\smallskip}\hline\noalign{\smallskip}
		1 & Phenanthrene	                    &   C$_{14}$H$_{10}$    &  20  & - $^4$ & 3.0087    &  --           \\
		2 & Chrysene	                        &   C$_{18}$H$_{12}$    &  33  & 5.4323 & 0.0896    &  60.6283     \\
		3 & Triphenylene	                    &   C$_{18}$H$_{12}$    &  50  & 6.5119 & -         &  --            \\
		4 & Benzo[a]anthracene	                &   C$_{18}$H$_{12}$    &  17  & -      & 3.3416    &  --            \\
		5 & Perylene	                        &   C$_{20}$H$_{12}$    &  33  & 4.4219 & 0.7326    &  6.0359        \\
		6 & Benzo[a]Pyrene	                    &   C$_{20}$H$_{12}$    &  17  & -      & 3.5903    &  --           \\
		7 & Benzo[e]Pyrene	                    &   C$_{20}$H$_{12}$    &  33  & -      & 0.8320    &  --            \\
		8 & Benzo-ghi-perylene	                &   C$_{22}$H$_{12}$    &  17  & -      & 3.8310    &  --            \\
		9 & PAH62	                            &   C$_{22}$H$_{14}$    &  43  & 3.3703 & 0.1518    &  22.2022     \\
		10 & Dibenzo-b-def-chrysene	            &   C$_{24}$H$_{14}$    &  29  & 6.5668 & 0.3759    &  17.4695     \\
		11 & PAH65	                            &   C$_{26}$H$_{14}$    &  43  & 4.7020 & 0.2967    &  15.8477     \\
		12 & Dibenzo-cd-lm-perylene	            &   C$_{26}$H$_{14}$    &  29  & 6.3810 & 9.5837    &  0.6658      \\
		13 & Bisanthene	                        &   C$_{28}$H$_{14}$    &  29  & -      & 3.9149    &  --          \\
		14 & Bezo-a-coronene	                &   C$_{28}$H$_{14}$    &  29  & 4.2570 & 7.1959    &  0.5916      \\
		15 & Dibenzo[fg,st]pentacene	        &   C$_{28}$H$_{16}$    &  50  & 4.4276 & 4.9596    &  0.8927      \\
		16 & Dibenzo-bc-ef-coronene	            &   C$_{30}$H$_{14}$    &  14  & -      & 6.8488    &  --          \\
		17 & Naphtho[8,1,2abc]coronene	        &   C$_{30}$H$_{14}$    &  14  & -      & 8.2805    &  --          \\
		18  & PAH66	                            &   C$_{30}$H$_{16}$    &  50  & 5.6046 & 0.2405    &  23.3040     \\
		19  & Terrylene	                        &   C$_{30}$H$_{16}$    &  50  & -      & 6.4576    &  --          \\
		20  & PAH67	                            &   C$_{34}$H$_{18}$    &  56  & 7.3400 & 0.4854    &  15.1215     \\
		21  & Tetrabenzocoronene	            &   C$_{36}$H$_{16}$    &  25  & -      & 4.5748    &  --          \\
		22  & benzo[a]ovalene                   &   C$_{36}$H$_{16}$    &  25  & -      & 8.8417    &  --          \\
		23  & Dibenzo[hi,yz]heptacene	        &   C$_{36}$H$_{20}$    &  40  & 1.1807 & 6.3096    &  0.1871      \\
		24  & Circumbiphenyl 	                &   C$_{38}$H$_{16}$    &  25  & 6.9933 & 11.1395   &  0.6278        \\
		25  & Naphth[8,2,1,abc]ovalene	        &   C$_{38}$H$_{16}$    &  13  & -      & 9.1260    &  --          \\
		26  & PAH64	                            &   C$_{38}$H$_{20}$    &  35  & 2.2953 & 1.9427    &  1.1815      \\
		27  & PAH68	                            &   C$_{38}$H$_{20}$    &  55  & 5.4204 & 0.3564    &  15.2088     \\
		28  & Phenanthro[3,4,5,6 vuabc]ovalene  &   C$_{40}$H$_{16}$    &  13  & -      & 10.9727   &  --           \\
		29  & Quaterrylene	                    &   C$_{40}$H$_{20}$    &  60  & 3.9990 & 2.5500    &  1.5682      \\
		30  & PAH69	                            &   C$_{40}$H$_{22}$    &  55  & 6.1832 & 0.2588    &  23.8918     \\
		31  & Dibenz[jk,a1b1]octacene	        &   C$_{40}$H$_{22}$    &  36  & 1.5360 & 8.4421    &  0.1819      \\
		32 & Hexabenzocoronene                  &   C$_{42}$H$_{18}$    &  67  & 5.1698 & 12.5668   &  0.4114       \\
		33 & Honeycomb15	                    &   C$_{46}$H$_{18}$    &  22  & -      & 9.4177    &  --          \\
		34 & Honeycomb16	                    &   C$_{48}$H$_{18}$    &  22  & -      & 8.3141    &  --          \\ 
		35 & Dicoronylene	                    &   C$_{48}$H$_{20}$    &  20  & 6.2168 & 12.2820   &  0.5062       \\
		36 & Kekulene	                        &   C$_{48}$H$_{24}$    &  25  & 2.1244 & 13.0511   &  0.1628      \\
		37 & Honeycomb17	                    &   C$_{50}$H$_{18}$    &  11  & -      & 17.8705   &  --           \\
		38 & Pentarylene	                    &   C$_{50}$H$_{24}$    &  67  & 4.3661 & 3.5769    &  1.2206      \\
		39 & PAH70	                            &   C$_{50}$H$_{26}$    &  50  & 4.7942 & 2.5430    &  1.8853      \\
		40 & PAH71	                            &   C$_{54}$H$_{28}$    &  57  & 5.1842 & 0.2294    &  22.5990     \\
		41 & Honeycomb19	                    &   C$_{56}$H$_{20}$    &  20  & -      & 10.1078   &  --           \\
		42 & Honeycomb20	                    &   C$_{58}$H$_{20}$    &  10  & -      & 8.8576    &  --          \\
		43 & Honeycomb21	                    &   C$_{62}$H$_{22}$    &  27  & -      & 4.9764    &  --          \\
		44 & Honeycomb22	                    &   C$_{64}$H$_{22}$    &  18  & -      & 4.6413    &  --          \\
		45 & Honeycomb23	                    &   C$_{66}$H$_{22}$    &  18  & -      & 11.2213   &  --           \\
		46 & Honeycomb24	                    &   C$_{70}$H$_{24}$    &  42  & -      & 8.6362    &  --          \\
		47 & Honeycomb25	                    &   C$_{72}$H$_{24}$    &  42  & 2.2637 & 7.4262    &  0.3048      \\
		48 & Honeycomb26	                    &   C$_{74}$H$_{24}$    &  42  & 2.3387 & 7.0688    &  0.3308      \\
		49 & Honeycomb27	                    &   C$_{78}$H$_{26}$    &  54  & 1.7217 & 0.2593    &  6.6398      \\
		50 & Honeycomb28	                    &   C$_{80}$H$_{26}$    &  54  & 1.7118 & 0.2573    &  6.6529      \\
		51 & Honeycomb29	                    &   C$_{82}$H$_{26}$    &  54  & 1.6768 & 0.2990    &  5.6080      \\
		52 & Honeycomb30	                    &   C$_{84}$H$_{26}$    &  54  & 2.4132 & 0.6710    &  3.5964      \\
		\noalign{\smallskip}\hline	
	    \footnote[1]{intrinsic band  strengths of pure bay/non-bay C-H stretching mode}
	    \footnote[2]{in units of km/mol/numbers of bay/nonbay C-H bonds in the molecular structure}
		\footnote[3]{Sum of the intensity of pure bay/nonbay transitions divided by the numbers of bay/nonbay C-H bonds in the molecular structure}
		\footnote[4]{No pure bay/nonbay transitions or found to be entirely Raman modes}					
	\end{longtable}

\clearpage

%\begin{table}[h]

%	\label{tab6:allyl-PAH}   
%	\begin{center} 
		\begin{longtable}{cccc}
\caption{Names and molecular formulas for  allyl-PAH molecules in Figure \ref{ally}} \label{allyl-PAH}	\\
	
			\hline\noalign{\smallskip}
			No & Name & Formula \\
			\noalign{\smallskip}\hline\noalign{\smallskip}
			
			53  & allylbenzene                                 & C$_{9}$H$_{10}$    \\
			54  & di-allylbenzene                               & C$_{12}$H$_{14}$   \\
			55  & tri-allylbenzene                              & C$_{15}$H$_{18}$   \\
			56  & allylnaphthalene                             & C$_{13}$H$_{12}$   \\
			57  & di-allylnaphthalene							 & C$_{16}$H$_{16}$   \\
			58  & tri-allylnaphthalene   						 & C$_{19}$H$_{20}$   \\
			59  & allyanthracene                               & C$_{17}$H$_{14}$   \\
			60  & di-allyanthracene							 & C$_{20}$H$_{18}$   \\
			61  & tri-allylanthracene                           & C$_{23}$H$_{22}$   \\
			62  & allylphenanthrene							 & C$_{17}$H$_{14}$   \\
			63  & di-allylphenanthrene							 & C$_{20}$H$_{18}$   \\
			64  & tri-allylphenanthrene						 & C$_{23}$H$_{22}$   \\
			65  & allylpyrene                                  & C$_{19}$H$_{14}$   \\
			66  & di-allylpyrene								 & C$_{22}$H$_{18}$   \\
			67  & tri-allylpyrene								 & C$_{25}$H$_{22}$   \\
			68  & allyltetracene								 & C$_{21}$H$_{16}$   \\
			69  & di-allyltetracene							 & C$_{24}$H$_{20}$   \\
			70  & tri-allyltetracene                            & C$_{27}$H$_{24}$   \\
			71  & allylchrysene								 & C$_{21}$H$_{16}$   \\
			72  & di-allylchrysene								 & C$_{24}$H$_{20}$   \\
			73  & tri-allylchrysene							 & C$_{27}$H$_{24}$   \\
			74  & allyltriphenylene							 & C$_{21}$H$_{16}$   \\
			75  & di-triphenylene								 & C$_{24}$H$_{20}$   \\
			76  & tri-triphenylene								 & C$_{27}$H$_{24}$   \\
			77  & allylbenzo-a-anthracene                      & C$_{21}$H$_{16}$   \\
			78  & di-allylbenzo-a-anthracene                    & C$_{24}$H$_{20}$   \\
			79  & tri-allylbenzo-a-anthracene 					 & C$_{27}$H$_{24}$   \\
			80  & allylperylene                                & C$_{23}$H$_{16}$   \\
			81  & di-allylperylene								 & C$_{26}$H$_{20}$   \\
			82  & tri-allylperylene							 & C$_{29}$H$_{24}$   \\
			83  & allylbenzo-a-pyrene							 & C$_{23}$H$_{16}$   \\
			84  & di-allylbenzo-a-pyrene						 & C$_{26}$H$_{20}$   \\
			85  & tri-allylbenzo-a-pyrene	                     & C$_{29}$H$_{24}$   \\
			86  & allylbenzo-e-pyrene                          & C$_{23}$H$_{16}$   \\
			87  & di-allylbenzo-e-pyrene                        & C$_{26}$H$_{20}$   \\
			88  & tri-allylbenzo-e-pyrene                       & C$_{29}$H$_{24}$   \\
			89  & allyanthanthrene							 & C$_{25}$H$_{16}$   \\
			90  & di-allyanthanthrene							 & C$_{28}$H$_{20}$   \\
			91  & tri-allyanthanthrene                          & C$_{31}$H$_{24}$   \\
			92  & allylbenzo-ghi-perylene                      & C$_{25}$H$_{16}$   \\
			93  & di-allylbenzo-ghi-perylene					 & C$_{28}$H$_{20}$   \\
			94  & tri-allylbenzo-ghi-perylene                   & C$_{31}$H$_{24}$   \\
			95  & allylpentacene                              & C$_{25}$H$_{18}$   \\
			96  & di-allylpentacene 							 & C$_{28}$H$_{22}$   \\
			97  & tri-allylpentacene 							 & C$_{31}$H$_{26}$   \\
			98  & allylcoronene								 & C$_{27}$H$_{16}$   \\
			99  & di-allylcoronene								 & C$_{30}$H$_{20}$   \\
			100  & tri-allylcoronene                             & C$_{33}$H$_{24}$   \\
			101  & allyldibenzo-bdef-chrysene                   & C$_{27}$H$_{18}$   \\
			102  & di-allyldibenzo-bdef-chrysene                 & C$_{30}$H$_{22}$   \\
			103  & tri-allyldibenzo-bdef-chrysene                & C$_{33}$H$_{26}$   \\
			104  & allyldibezo-cdlm-perylene                    & C$_{29}$H$_{18}$   \\
			105  & di-allyldibezo-cdlm-perylene                  & C$_{32}$H$_{22}$   \\
			106  & tri-allyldibezo-cdlm-perylene				 & C$_{35}$H$_{26}$   \\
			107  & allylhexacene                                & C$_{29}$H$_{20}$   \\
			108  & di-allylhexacene                              & C$_{32}$H$_{24}$   \\
			109  & tri-allylhexacene                             & C$_{35}$H$_{28}$   \\
			110  & allylbisanthene                              & C$_{31}$H$_{18}$   \\
			111  & di-allylbisanthene                            & C$_{34}$H$_{22}$   \\
			112  & tri-allylbisanthene                           & C$_{35}$H$_{28}$   \\
			113  & allylovalene                                 & C$_{35}$H$_{18}$   \\
			114  & hepta-allylovalene                          & C$_{53}$H$_{42}$   \\
			
			\noalign{\smallskip}\hline
%		\end{tabular}
%	\end{center}
\end{longtable}

\begin{table}[h]
	\caption{Names and molecular formulas for  alkyl-PAH molecules in Figure \ref{alkyl}}
	\label{alkyl-PAH}   
	\begin{center} 
		\begin{tabular}{llll}
			\hline\noalign{\smallskip}
			No & Name & Formula \\
			\noalign{\smallskip}\hline\noalign{\smallskip}
	
	115     &   ethylbenzene					 	&	C$_{8}$H$_{10}$   \\
	116    &   propylnathelene				 		&	C$_{13}$H$_{14}$  \\
	117    &   butylanthracene				 		&	C$_{18}$H$_{18}$  \\
	118     &   butylphenanthrene				 	&	C$_{18}$H$_{18}$  \\
	119     &   butylpyrene					 		&	C$_{20}$H$_{18}$  \\
	120     &   pentyltetracene				 		&	C$_{23}$H$_{22}$  \\
	121     &   pentylchrysene					 	&	C$_{23}$H$_{22}$  \\
	122     &   pentyltriphenylene				 	&	C$_{23}$H$_{22}$  \\
	123     &   pentylbenzo-a-anthracene		 	&	C$_{23}$H$_{22}$  \\
	124     &   pentylperylene					 	&	C$_{25}$H$_{22}$  \\
	125     &   pentylbenzo-a-pyrene			 	&	C$_{25}$H$_{22}$  \\
	126     &   pentylbenzo-e-pyrene			 	&	C$_{25}$H$_{22}$  \\
	127     &   pentylanthanthrene				 	&	C$_{27}$H$_{22}$  \\
	128     &   pentylbenzo-ghi-perylene		 	&	C$_{27}$H$_{22}$  \\
	129     &   hexylpentacene					 	&	C$_{28}$H$_{26}$  \\
	130     &   pentylcoronen					 	&	C$_{29}$H$_{22}$  \\
	131     &   hexyl-dibenzo-bdef-chrysene	 		&	C$_{30}$H$_{26}$  \\
	132     &   hexyl-dibenzo-cdlm-perylene	 		&	C$_{32}$H$_{26}$  \\
	133     &   heptylhexacene					 	&	C$_{33}$H$_{30}$  \\
	134     &   hexylbisanthene				 		&	C$_{34}$H$_{26}$  \\
	135     &   dodecylHoneycomb30				 	&	C$_{96}$H$_{50}$  \\		
			
\noalign{\smallskip}\hline
		\end{tabular}
	\end{center}
\end{table}
			
\clearpage

\begin{table}[h]
	\caption{Names and molecular formulas for aliphatic molecules in Figure \ref{alkane}}
	\label{aliphatics}   
	\begin{center} 
		\begin{tabular}{llll}
			\hline\noalign{\smallskip}
			No & Name & Formula \\
			\noalign{\smallskip}\hline\noalign{\smallskip}
			
			136     &   ethane	                    &       C$_{2}$H$_{6}$      \\
			137     &   n-propane				    &	    C$_{3}$H$_{8}$      \\
			138    &   n-butane				    &	    C$_{4}$H$_{10}$     \\
			139     &   n-pentane				    &	    C$_{5}$H$_{12}$     \\
			140     &   n-hexane				    &	    C$_{6}$H$_{14}$     \\
			141     &   n-heptane				    &	    C$_{7}$H$_{16}$     \\
			142     &   n-octane				    &	    C$_{8}$H$_{18}$     \\
			143     &   n-nonane				    &	    C$_{9}$H$_{20}$     \\
			144     &   n-decane				    &	    C$_{10}$H$_{22}$    \\
			145     &   n-icosane				    &	    C$_{20}$H$_{42}$    \\
			146     &   n-triacontane			    &	    C$_{30}$H$_{62}$    \\
			147     &   n-tetracontane			    &	    C$_{40}$H$_{82}$    \\
			148     &   n-pentacontane			    &	    C$_{50}$H$_{102}$   \\
			149     &   2,2-dimethylbutane		    &	    C$_{6}$H$_{14}$     \\
			150     &   3,3-dimethyloctane		    &	    C$_{10}$H$_{22}$    \\
			151     &   hydrocarbon				    &	C$_{15}$H$_{32}$    \\
			152     &   hydrocarbon				    &	C$_{23}$H$_{48}$    \\
			153     &   hydrocarbon					&	C$_{31}$H$_{64}$    \\
			
			\noalign{\smallskip}\hline
		\end{tabular}
	\end{center}
\end{table}

\clearpage

\begin{table}[h]
	\caption{\label{DFT-PAH} Comparison between theory and experiment of wavelengths and intensities of 3 $\mu$m  vibrational modes  of neutral PAH molecules}
	\begin{center}
     \begin{tabular}{llll}
      {Method}&{Number of transitions}&{$\lambda $($\mu$m)}& {I(km.mol$^{-1}$)}\\
                Experiment$^a$ & 77 &3.28784&64.0916\\
		Theoretical (DFT)$^b$&1026&3.25468&10.9139\\
				Exp-DFT&-&0.03316&53.1777\\
			\end{tabular}
	\end{center}
	{$^a$}{\footnotesize 28 PAH molecules from NASA-AMES PAH database v2.0.}
	{$^b$}{\footnotesize 60 PAH molecules  \citep{SKZ2015-2}.}
\end{table}

\clearpage

\begin{figure}
	\begin{center}
		\resizebox{175mm}{!}{\includegraphics{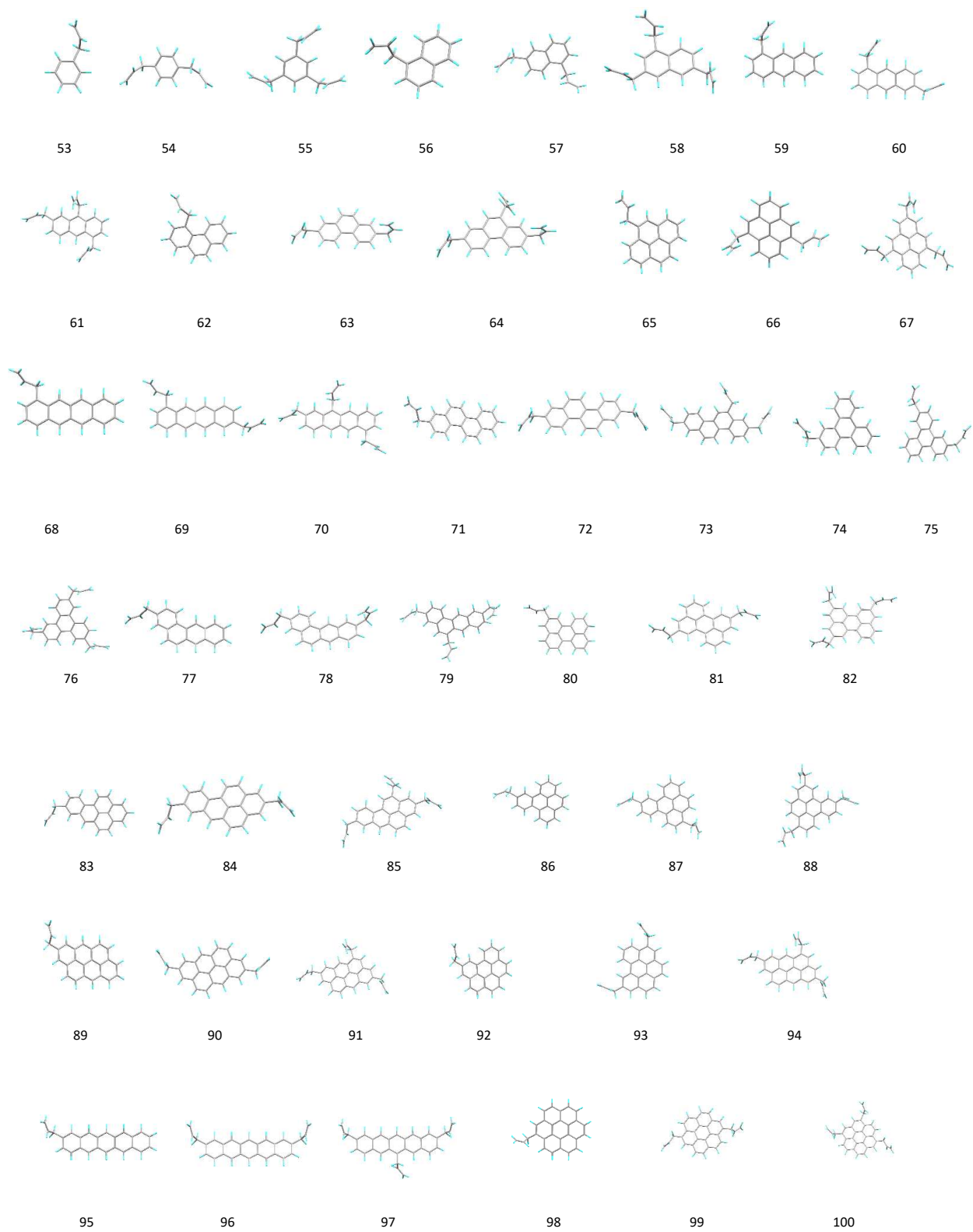}}
	\end{center}
	\caption{\label{ally} Local minimum geometries of 62 neutral allyl-PAH molecules.}		
\end{figure}

\begin{figure}
\setcounter{figure}{1}
	\begin{center}
		\resizebox{190mm}{!}{\includegraphics{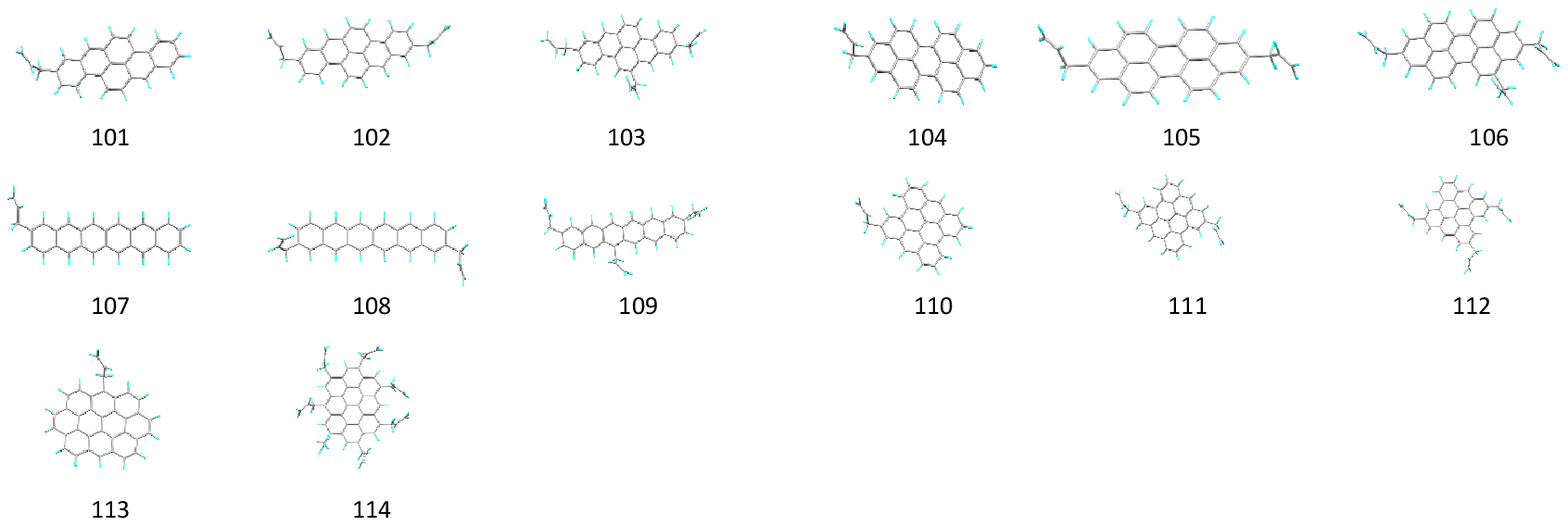}}
	\end{center}
\caption{Continued.}
\end{figure}

\clearpage
\begin{figure}
	\begin{center}
		\resizebox{190mm}{!}{\includegraphics{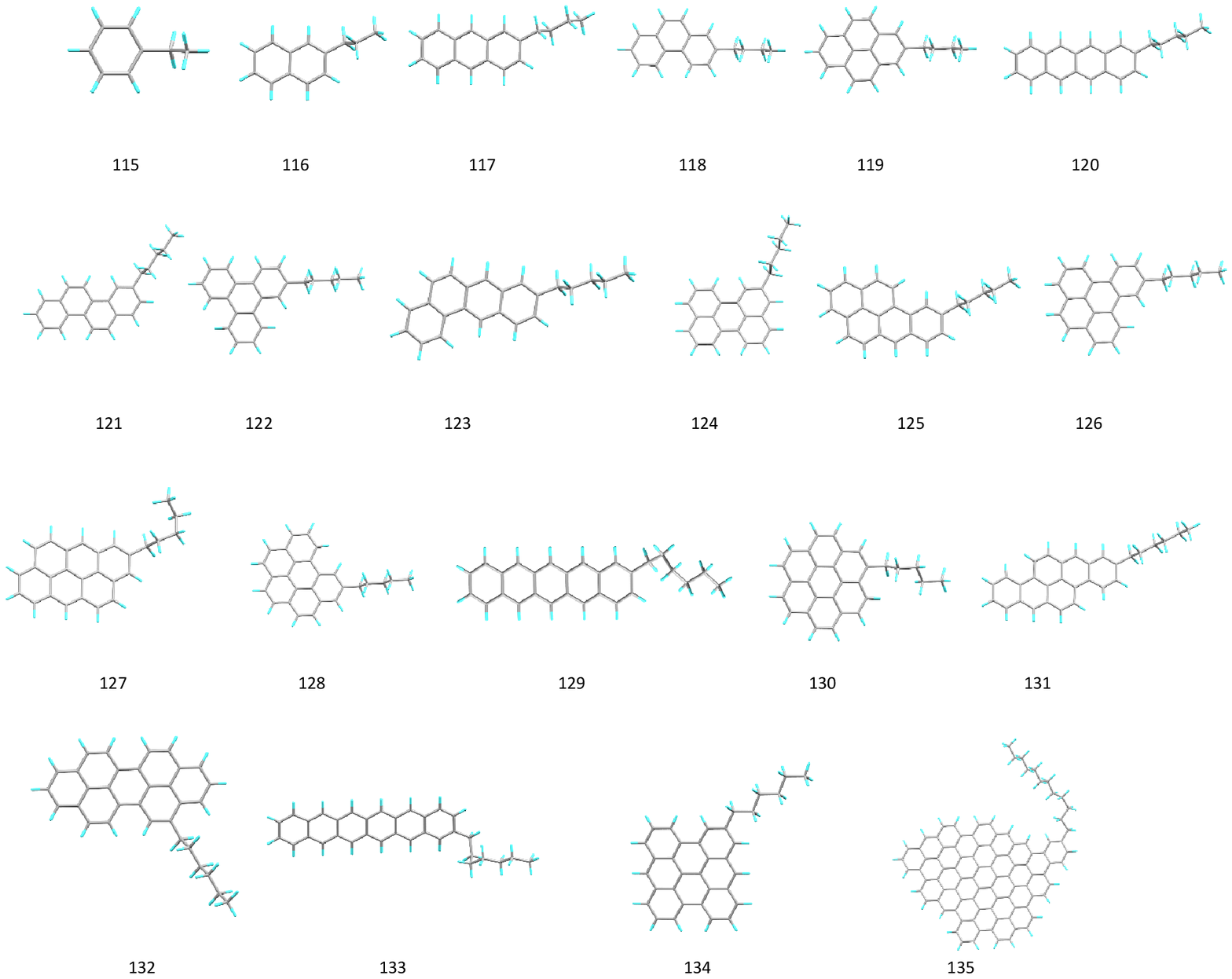}}
	\end{center}
	\caption{\label{alkyl} Local minimum geometries of 21 neutral alkyl-PAH molecules.  All these molecules are chosen to have equal number of aromatic and aliphatic C$-$H bonds}
	
\end{figure}

\clearpage

\begin{figure}
	\begin{center}
		\resizebox{190mm}{!}{\includegraphics{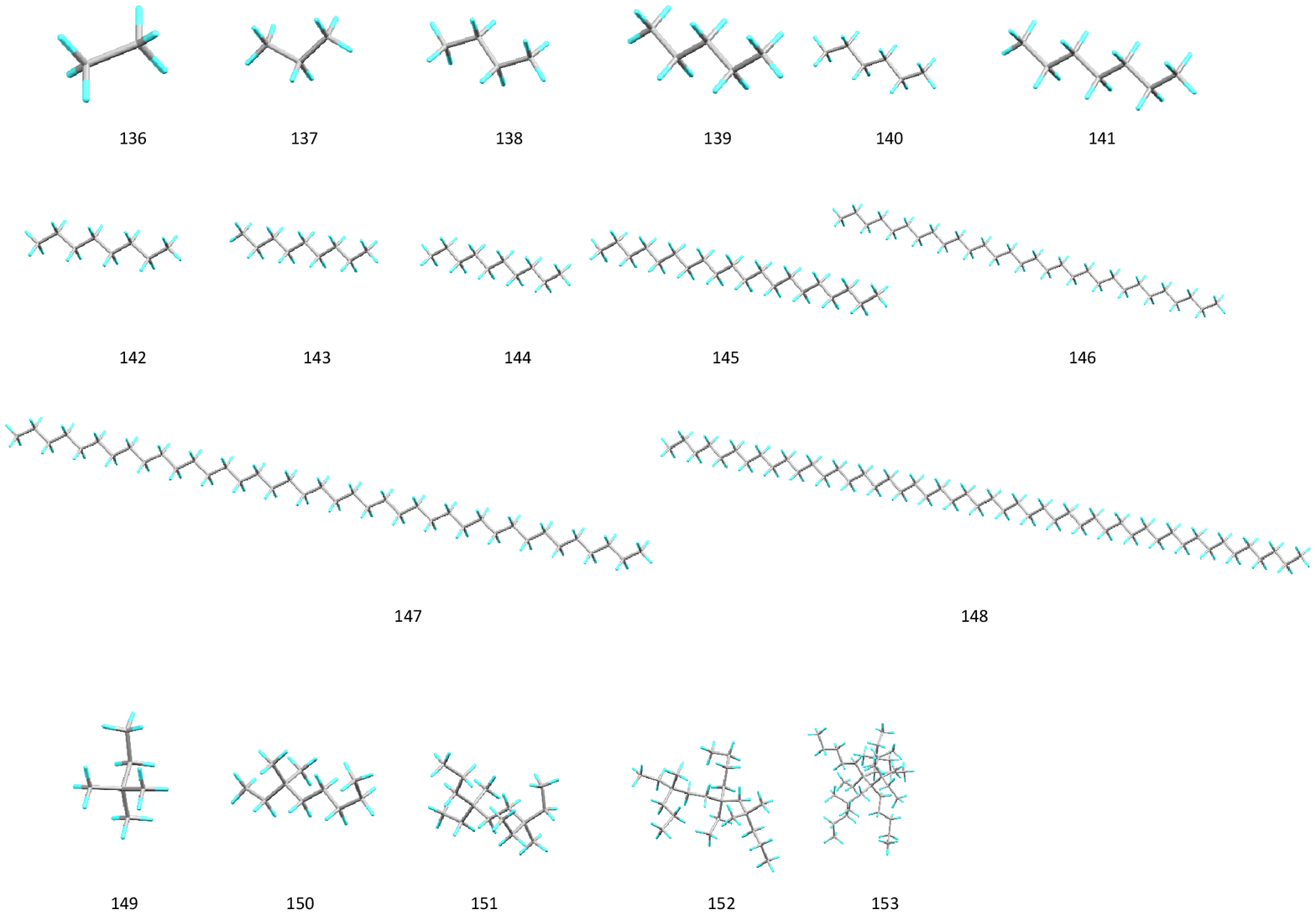}}
	\end{center}
	\caption{\label{alkane} Local minimum geometries of  18 molecules with linear and branched aliphatic $sp^3$ structures.}
\end{figure}

\clearpage

\begin{figure}
	\begin{center}
		\includegraphics[width=30pc]{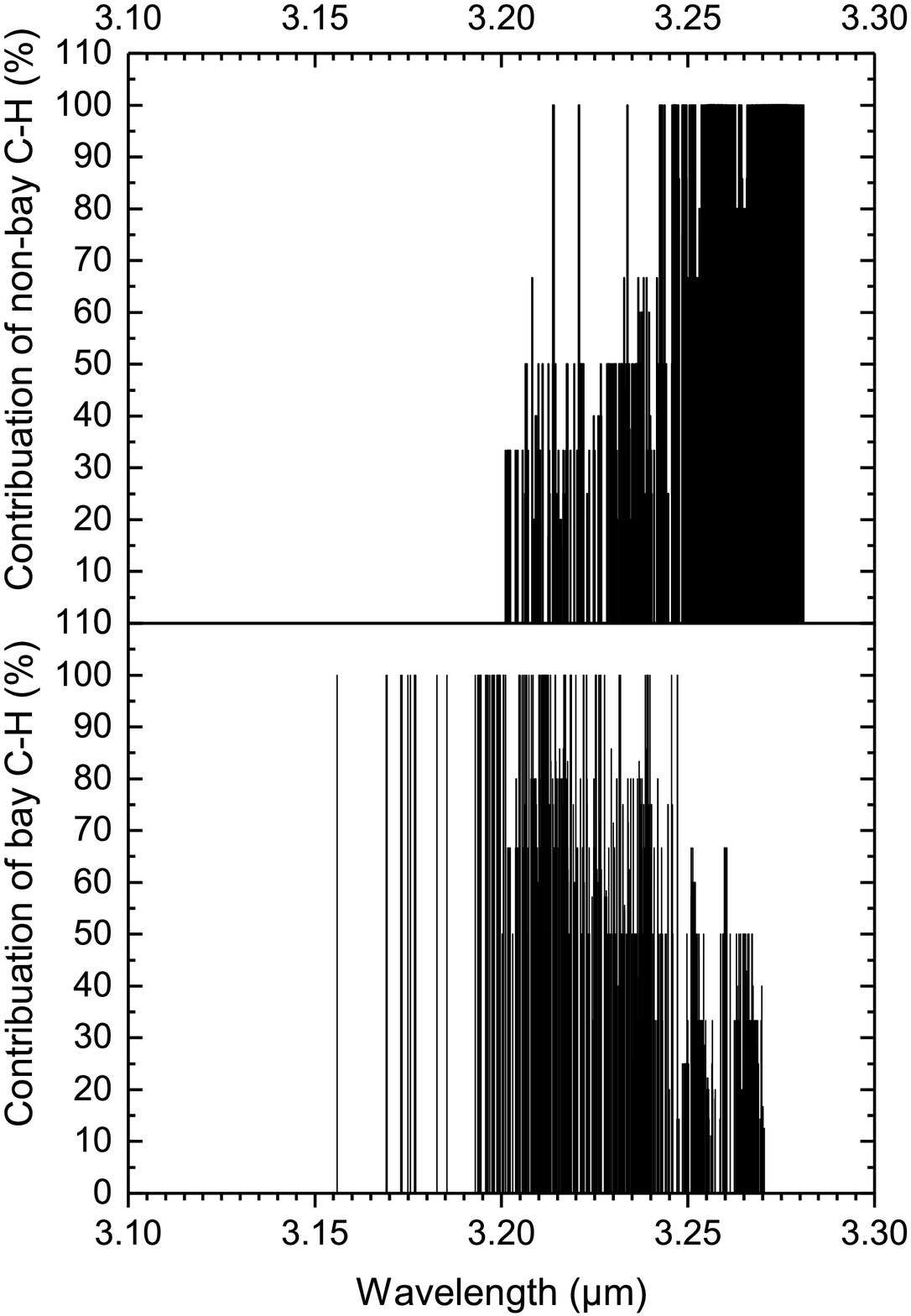}
	\end{center}
	\caption{\label{bay} Peak vibrational wavelengths dependence on the fraction  of `non-bay' (top panel) and `bay' (bottom panel) aromatic C$-$H stretching modes (in percentage scale) for the molecules  1--52 in Figure \ref{pah} and Table \ref{bay-PAH}).  The lines that have contributions less than 100\% are coupled modes.  Out of a total of 986 transitions plotted, 111 are pure `bay', 372 are pure `non-bay', and 503 are coupled modes.}
\end{figure}

\clearpage

\begin{figure}
	\begin{center}
		\includegraphics[width=30pc]{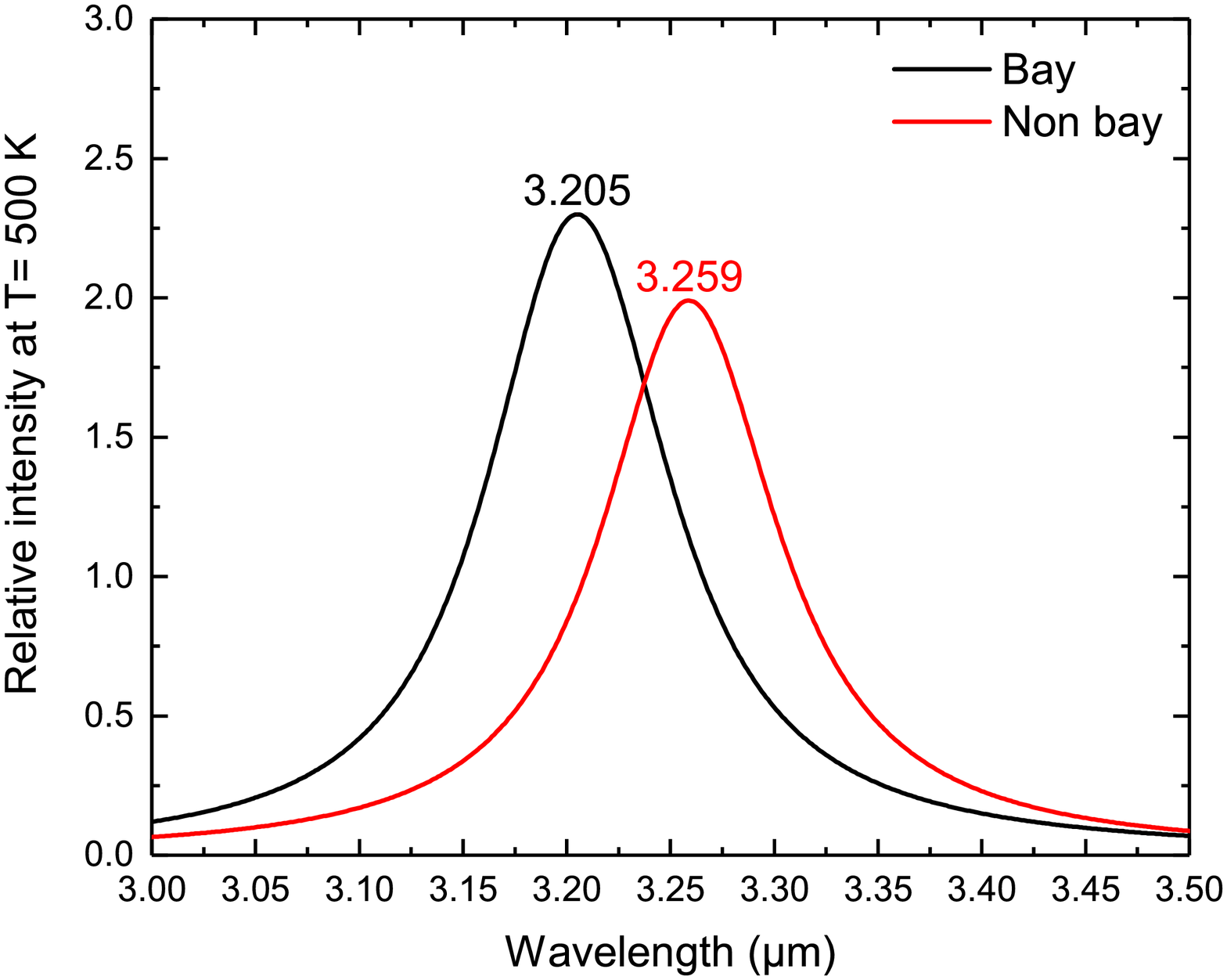}
	\end{center}
	\caption{\label{drude1} The simulated Drude profiles ($T$=500 K) of pure bay (111 transitions) and pure non-bay (372 transitions) aromatic C$-$H transitions.}
\end{figure}

\clearpage

\begin{figure}
	\begin{center}
		\includegraphics[width=30pc]{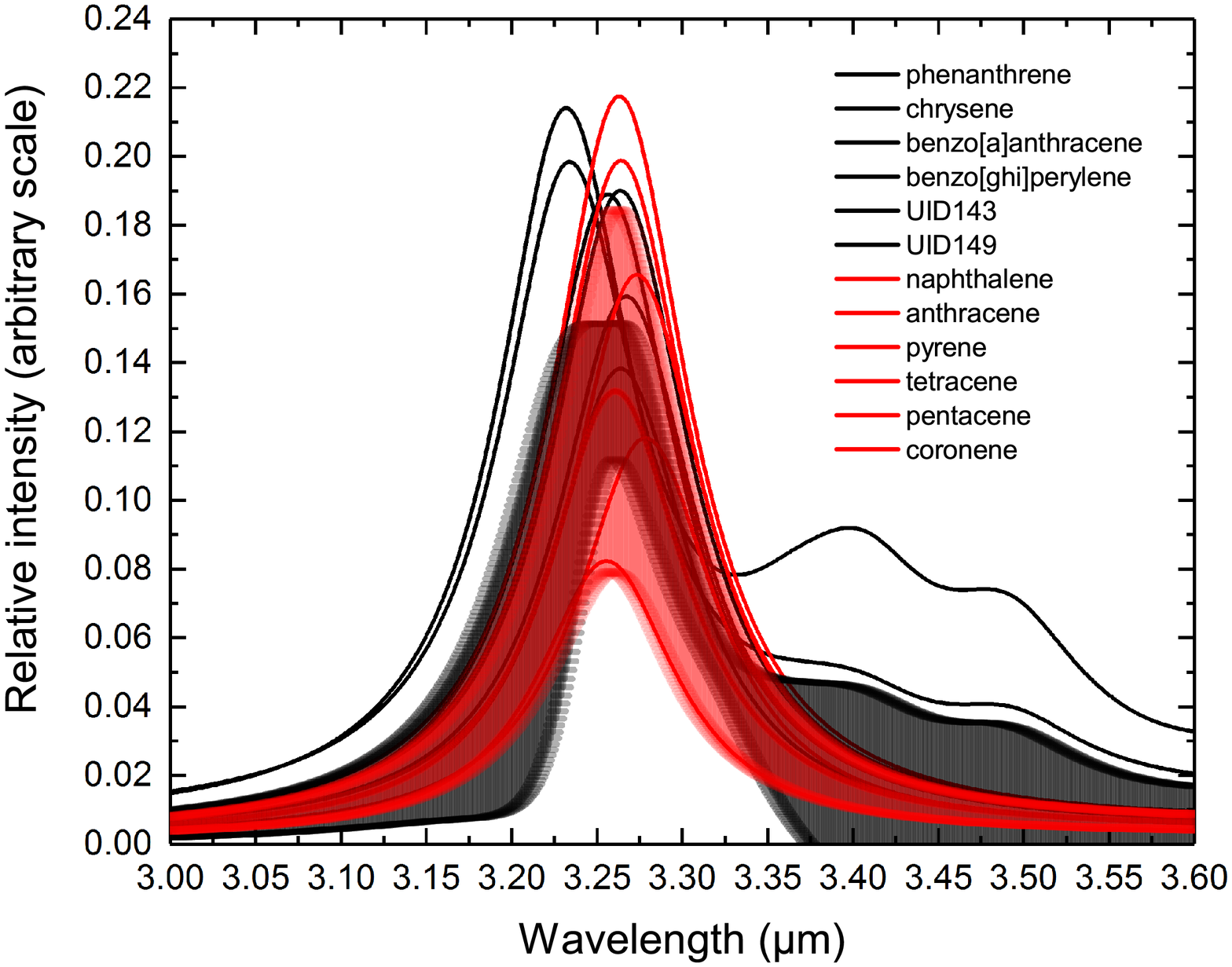}
	\end{center}
	\caption{\label{lab} Gas-phase laboratory infrared spectra of 12 neutral PAH molecules broadened by Drude profiles of  $T$=500 K and FWHM=0.03.  Bands profile of `bay' PAH molecules in in black and `non-bay' PAH molecules in red. The standard deviations of average spectra for group of `bay' and `non bay' PAH molecules are presented as shaded areas in black and red receptively. Data from NASA-PAH database v2.0.}
\end{figure}

\clearpage

\begin{figure}
	\begin{center}
		\includegraphics[width=30pc]{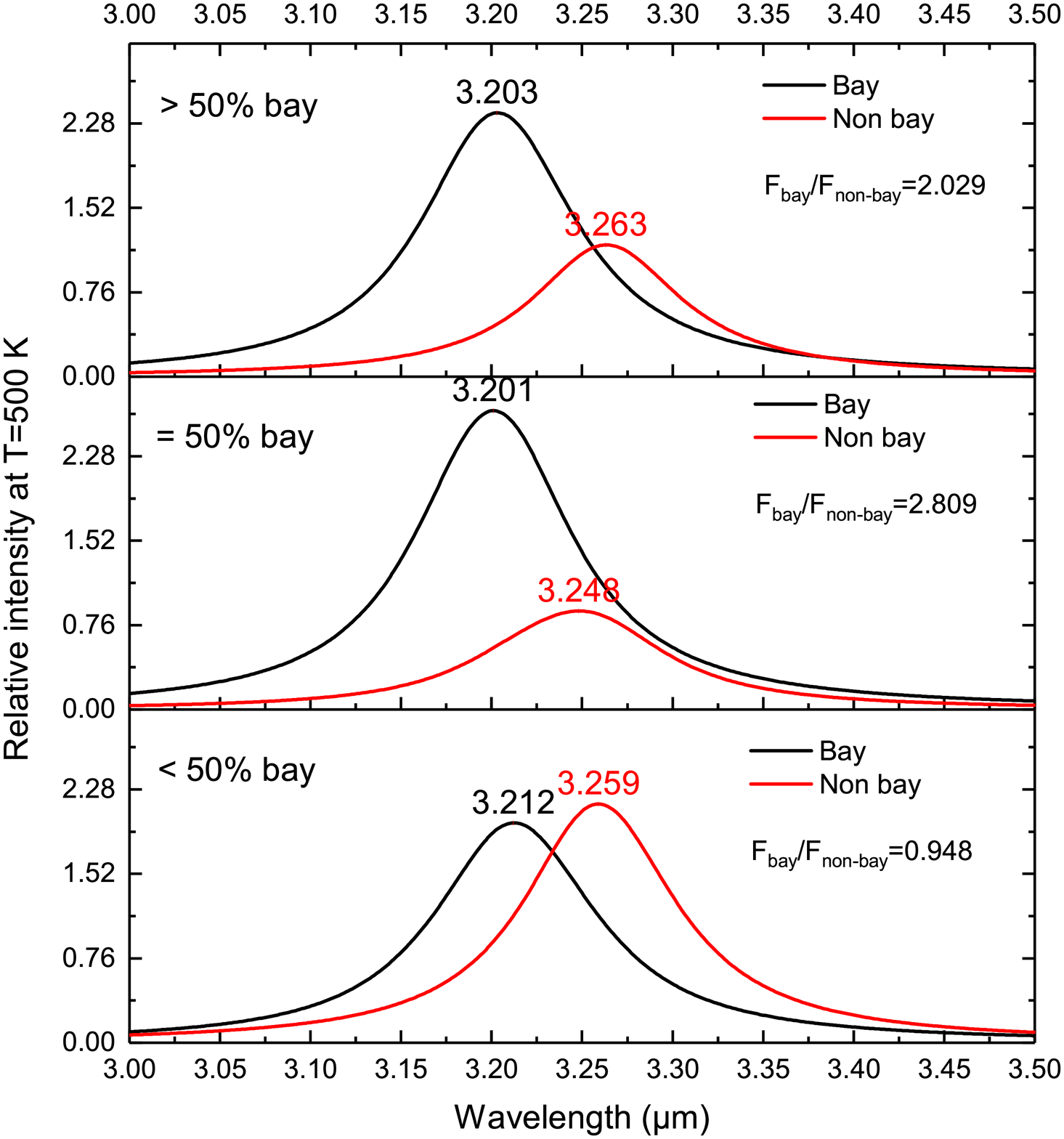}
	\end{center}
	\caption{\label{scheme1} Simulated astronomical 3.3 $\mu$m spectra for three groups of PAH molecules with different fractions of bay C--H bonds.  The vibrational modes plotted are all un-coupled modes.}
	
\end{figure}

\clearpage

%\begin{figure}
%	\begin{center}
%		\includegraphics[width=30pc]{fig13.eps}
%	\end{center}
%	\caption{\label{scheme2} The Drude simulated bands at $T$=500 K for PAH molecules classified based on scheme2. C$-$H transitions chosen here are all un-coupled modes.}
%	
%\end{figure}

%\clearpage

\begin{figure}
	\begin{center}
		\includegraphics[width=30pc]{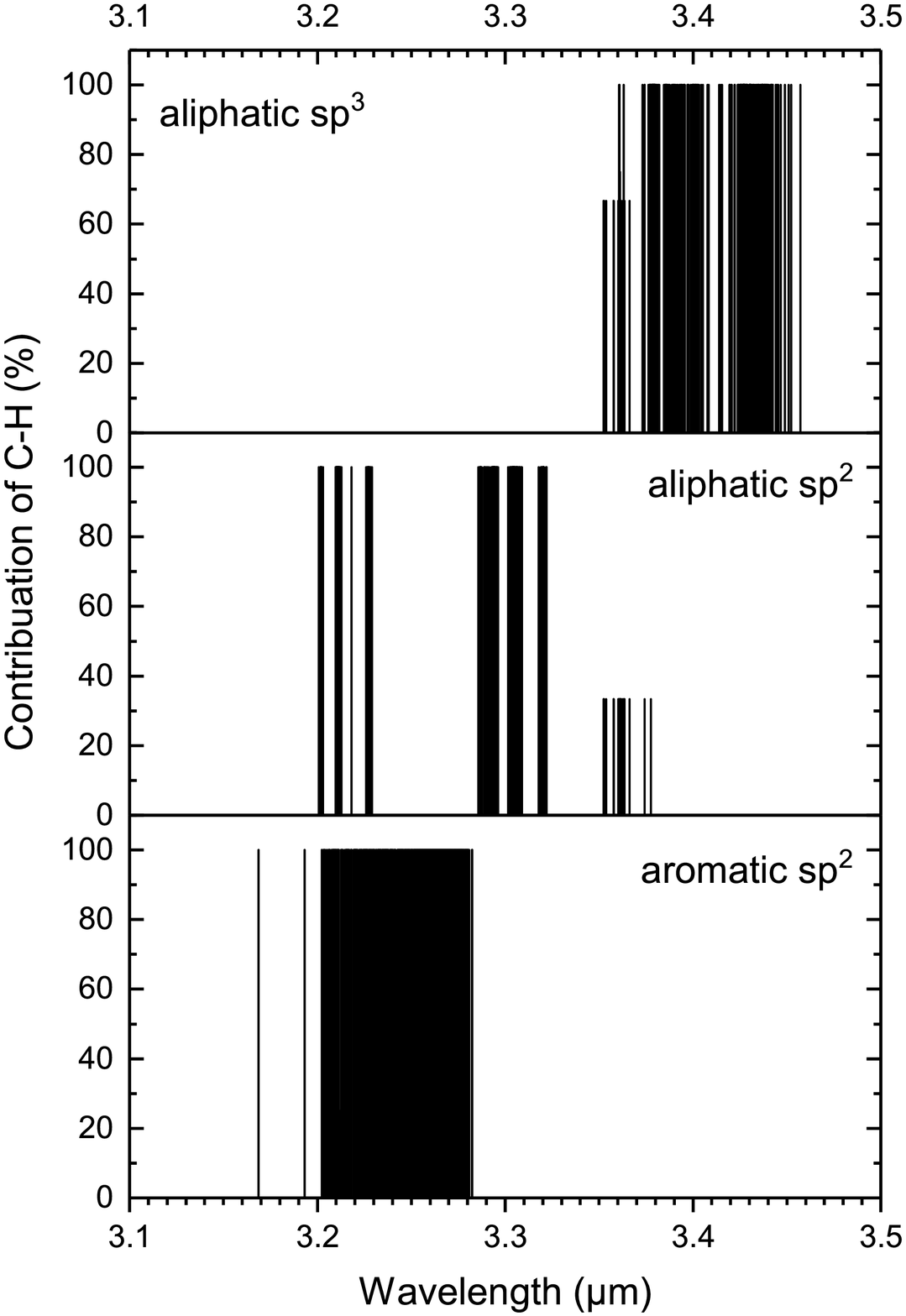}
	\end{center}
	\caption{\label{contributions} The contributions of aliphatic $sp^{3}$ (top panel), aliphatic $sp^{2}$ (olefinic) (middle panel), and  aromatic (bottom panel) C$-$H stretching modes in the 3 $\mu$m region for molecules 53--114 in Figure \ref{ally} and Table \ref{allyl-PAH}.}
\end{figure}

\clearpage

\begin{figure}
	\begin{center}
		\includegraphics[width=30pc]{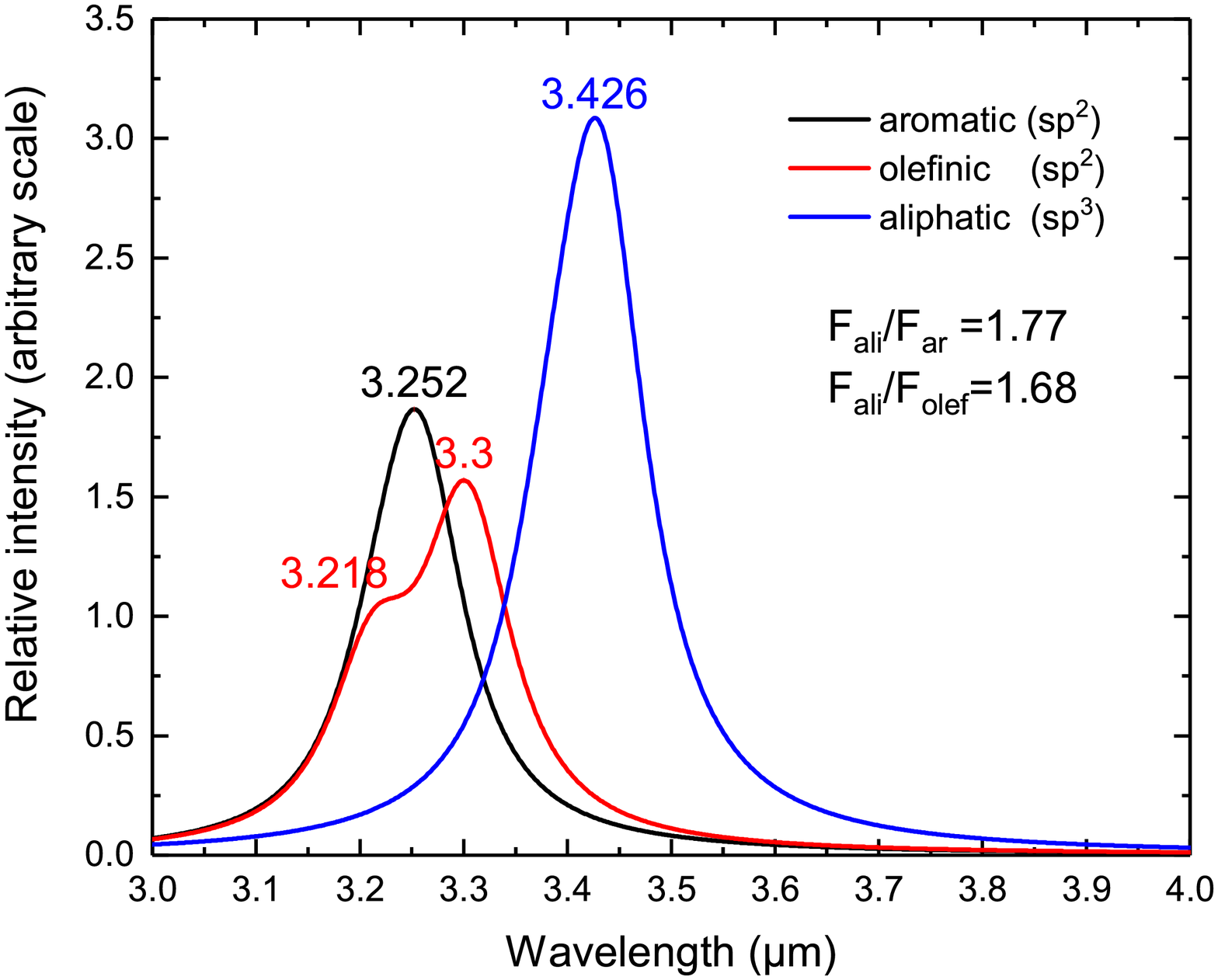}
	\end{center}
	\caption{\label{pure} Simulated astronomical spectra in the 3 $\mu$m region from purely aromatic, aliphatic $sp^{2}$ (olefinic), and aliphatic $sp^{3}$ C$-$H stretching modes in molecules with mixed aromatic/aliphatic structures (molecules 53--114 in Figure \ref{ally} and Table \ref{allyl-PAH}).  A Drude profile of $T$=500 K has been applied to the theoretical data.}
\end{figure}

\clearpage

\begin{figure}
	\begin{center}
		\includegraphics[width=30pc]{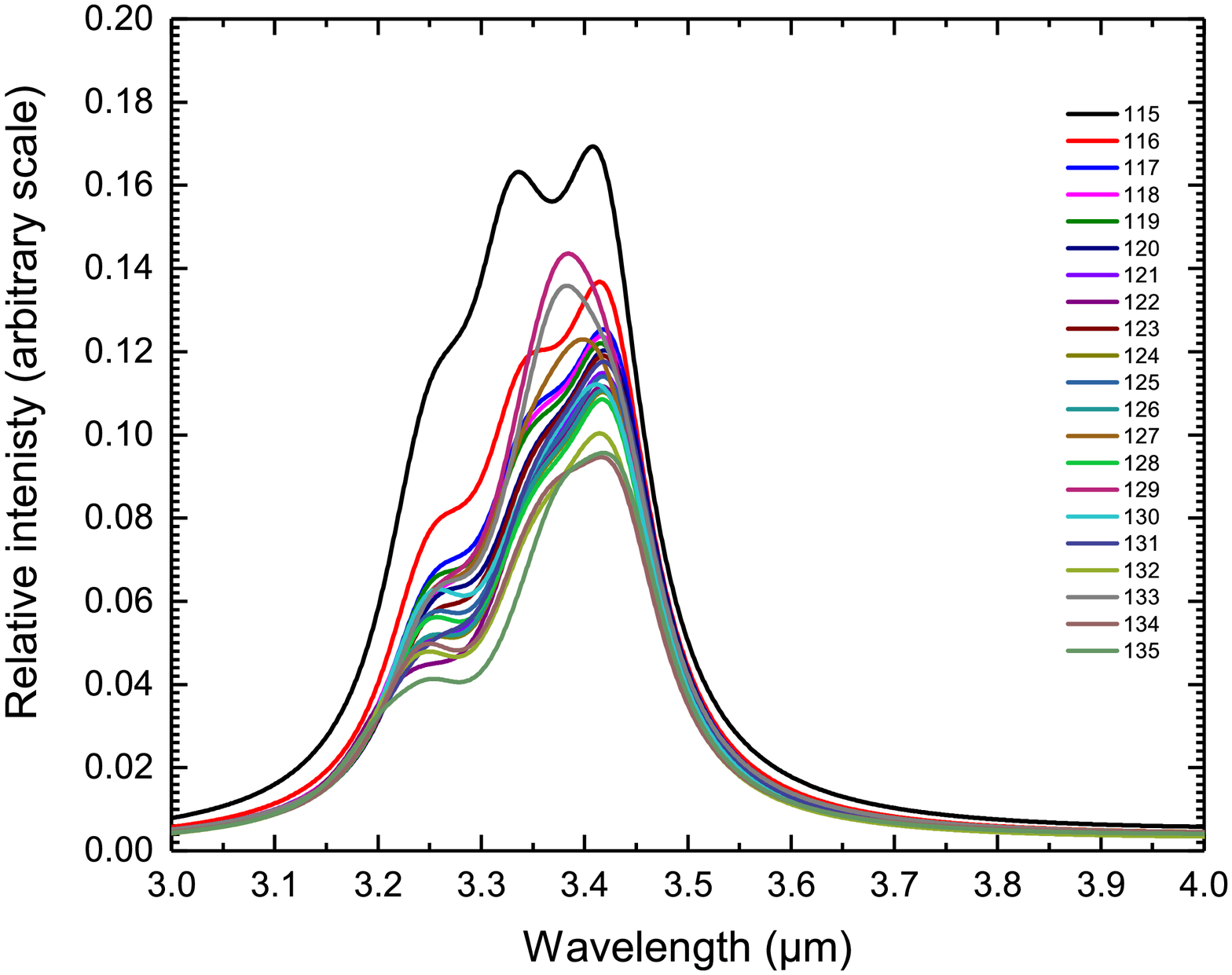}
	\end{center}
	\caption{\label{drude_alkyl} Simulated astronomical spectra in the  3 $\mu$m region for alkyl-aromatic molecules (structures 115--135 in Figure \ref{alkyl} and Table \ref{alkyl-PAH}).  A Drude profile of $T$=500 K is applied to the theoretical data.}
\end{figure}

\clearpage

%\begin{figure}
%	\begin{center}
%		\includegraphics[width=30pc]{fig18.eps}
%	\end{center}
%	\caption{\label{ethyl} The laboratory gas phase IR spectrum of ethylbenzene (C$_{6}$H$_{5}$CH$_{2}$CH$_{3}$).The integrated area under each band is in parenthesis in italic format.Data from NIST.} 
%\end{figure}
%
%\clearpage

\begin{figure}
	\begin{center}
		\includegraphics[width=30pc]{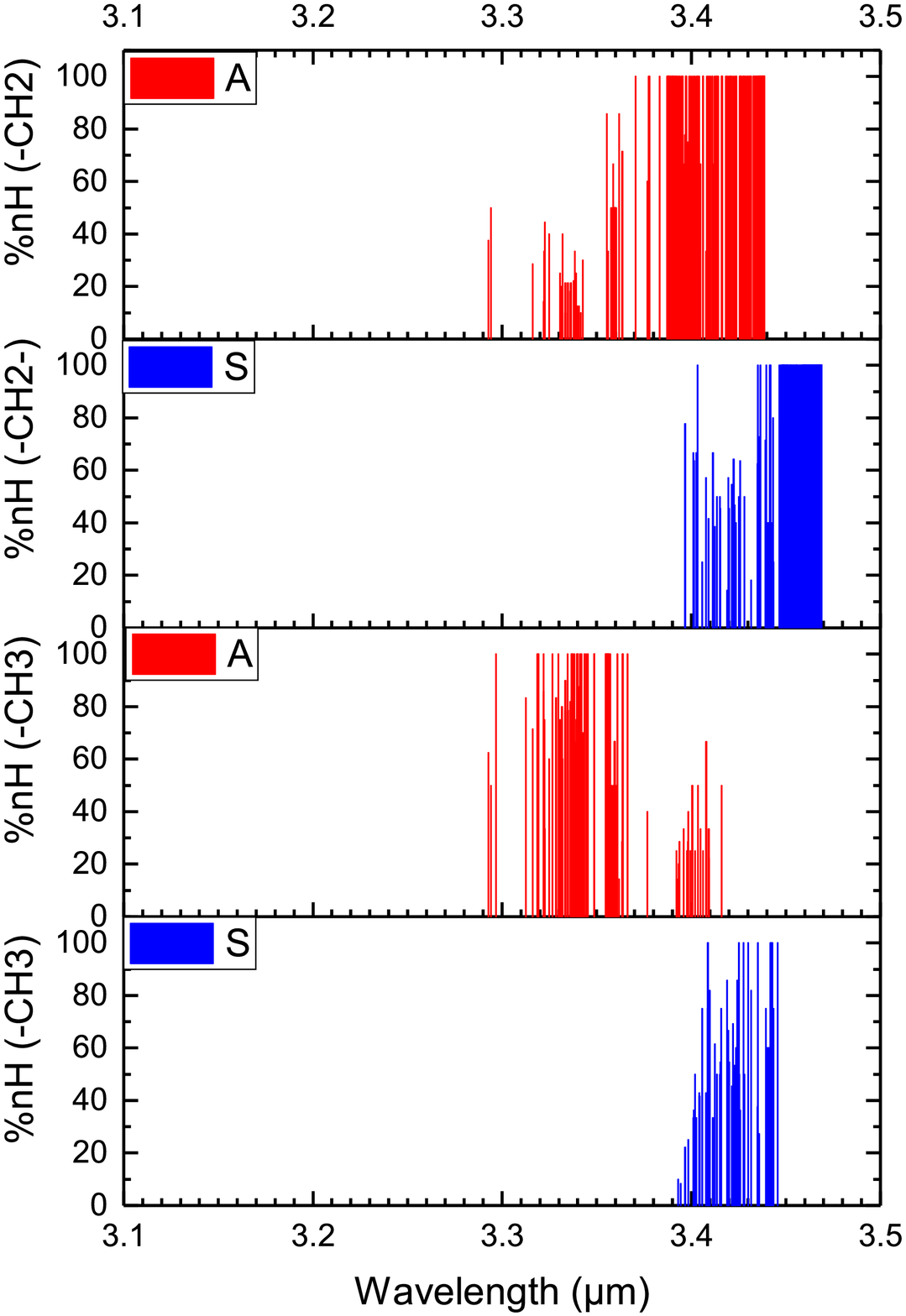}
	\end{center}
	\caption{\label{symm} Contributions of symmetric (S, in blue) and anti-symmetric (A, in red) of methyl (lower two panels) and methyelene (upper two panels) C--H stretching vibrations for molecules 136--153 in Figure \ref{alkane} and Table \ref{aliphatics}.}
	
\end{figure}

\clearpage

\begin{figure}
	\begin{center}
		\includegraphics[width=30pc]{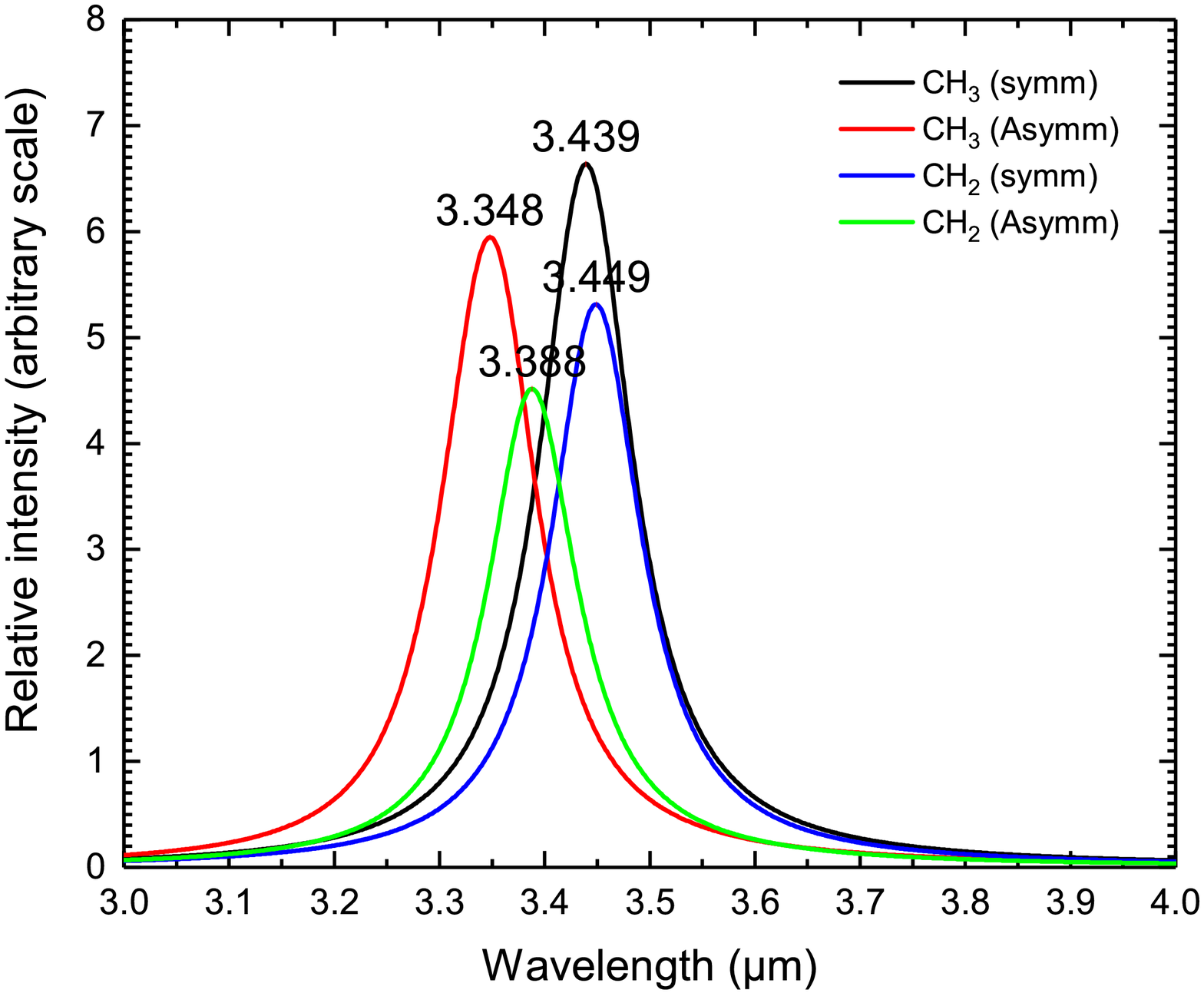}
	\end{center}
	\caption{\label{uncoupled}  Simulated astronomical spectra of  un-coupled  methyl and methylene C$-$H stretching modes in the 3 $\mu$m region for molecules 136--153 in Figure \ref{alkane} and Table \ref{aliphatics}.  A Drude profile of $T$=500 K is applied to the theoretical data.}
\end{figure}

\clearpage

\begin{figure}
	\begin{center}
		\includegraphics[width=30pc]{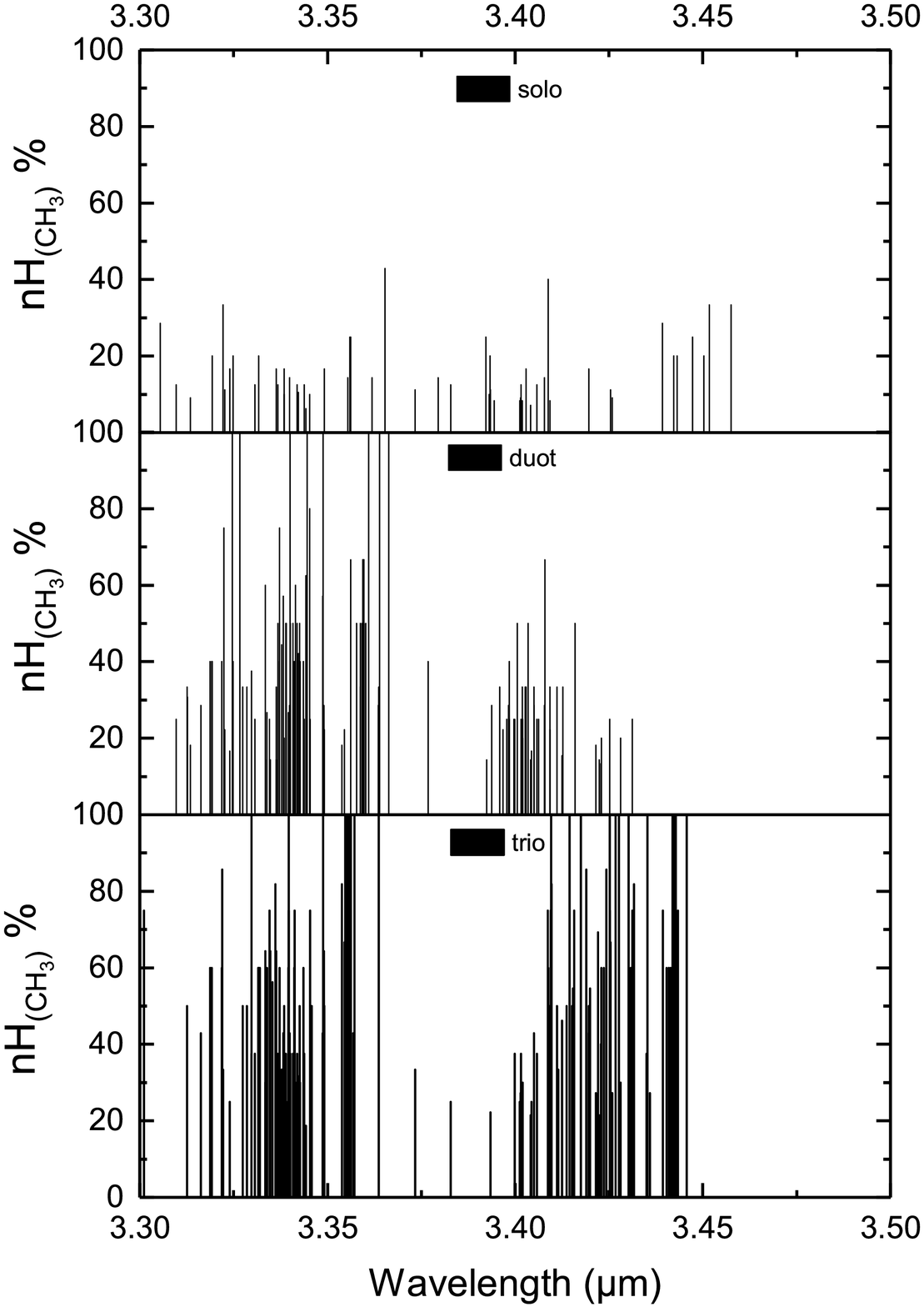}
	\end{center}
	\caption{\label{ch3} Classification and contributions of different types of vibrational motions of methyl groups in the 3 $\mu$m region for molecules 136--153 in Figure \ref{alkane} and Table \ref{aliphatics}.}
	
\end{figure}

\end{CJK*}


\clearpage

\begin{thebibliography}{10}






\bibitem[Allamandola, Tielens, \& Barker(1989)]{allamandola1989} 
Allamandola, L. J., Tielens, A. G. G. M., \& Barker, J. R. 1989, \apjs, 71, 733
	

\bibitem[Barker et al.(1987)]{barker} 
Barker, J.~R., Allamandola, L.~J., \& Tielens, A.~G.~G.~M.\ 1987, \apjl, 315, L61
	
	
\bibitem[Becke(1993a)]{becke1993a}  
Becke, A. D. 1993a, Journal of Chemical Physics, 98, 5648
	
\bibitem[Becke(1993b)]{becke1993b} 
Becke, A. D. 1993b, Journal of Chemical Physics, 98, 1372
	


\bibitem[Boersma et al.(2014)]{boersma} 
Boersma, C., Bauschlicher, C.~W., Jr., Ricca, A., et al.\ 2014, \apjs, 211, 8 

%	
\bibitem[Candian et al.(2012)]{Candian12} 
Candian, A., Kerr, T. H., Song, I.-O., McCombie, J., \& Sarre, P. J. 2012, \mnras, 426, 389

\bibitem[Candian et al.(2014)]{can14} 
Candian, A., Sarre, P.~J., \& Tielens, A.~G.~G.~M.\ 2014, \apjl, 791, L10 


	
\bibitem[Chiar et al.(2013)]{Chiar13}
Chiar, J. E., Tielens, A. G. G. M., Adamson, A. J., \& Ricca, A. 2013, \apj, 770, 78
		
%\bibitem[Clayton et al.(2011)]{Clayton11} 
%Clayton, G. C., De Marco, O., Whitney, B. A., et al. 2011, \aj, 142, 54
	
\bibitem[Coblentz(1905)]{Coblentz1905} 
Coblentz, William W. 1905 {\em Investigations of infra-red spectra} (Washington, D.C.: Carnegie Institution of Washington)
	
\bibitem[Colthup(1990)]{Colthup1990}
Colthup, N. B., Daly, L. H., \& Wiberley, S. E. (ed.) 1990, {\it Introduction to Infrared and Raman Spectroscopy} (Boston: Academic)


\bibitem[Cook et al.(1996)]{cook1996}
Cook, D. J., Schlemmer, S., Balucani, N., et al. 1996, \nat, 380, 227

\bibitem[Cook et al.(1998)]{cook1998}
Cook, D. J., Schlemmer, S., Balucani, N., et al. 1998, Journal of Physical Chemistry A, 102, 1465

\bibitem[Duley, \& Williams(1979)]{duley1979} 
Duley, W. W., \& Williams, D. A. 1979, \nat, 277, 40


\bibitem[Duley, \& Williams(1981)]{duley1981} 
Duley, W. W., \& Williams, D. A. 1981, \mnras, 196, 269
	
%\bibitem[Feller et al.(2010)]{Feller2010}
%Feller, David., Peterson, Kirk. A., \& Hill, J. Grant. 2010, JCP, 133, 184102
	

\bibitem[Fox \& Martin(1937)]{Martin1937} 
Fox, J. J., Martin, A. E. 1937, Proceedings of the Royal Society of London A: Mathematical, Physical and Engineering Sciences, 16, 419
	
\bibitem[Fox \& Martin(1938)]{Martin1938} 
Fox, J. J., Martin, A. E. 1938, Proceedings of the Royal Society of London A: Mathematical, Physical and Engineering Sciences, 167, 257
	
\bibitem[Fox \& Martin(1939)]{Martin1939} 
Fox, J. J., Martin, A. E. 1939, Journal of the Chemical Society (Resumed) 318
	
\bibitem[Fox \& Martin(1940)]{Martin1940} 
Fox, J. J., Martin, A. E. 1940, Proceedings of the Royal Society of London A: Mathematical, Physical and Engineering Sciences, 175, 208
	
\bibitem[Frisch et al.(2009)]{fri09}
Frisch, M. J., Frisch, G. W., Trucks, H. B., et al. 2009, Gaussian 09, Revision C.01, Gaussian, Inc., 
	


\bibitem[Geballe et al.(1992)]{geballe1992} 
Geballe, T. R., Tielens, A. G. G. M., Kwok, S., \& Hrivnak, B. J. 1992, \apj, 387, L89

\bibitem[Guillois, Ledoux, \& Reynaud(1999)]{guillois1999} 
Guillois, O., Ledoux, G., \& Reynaud, C. 1999, \apj, 521, L133	

\bibitem[Goto et al.(2003)]{goto03} 
Goto, M., Gaessler, W., Hayano, Y., et al.\ 2003, \apj, 589, 419


	
\bibitem[Goto et al.(2007)]{Goto07} 
Goto, M., Kwok, S., Takami, H., et al. 2007, \apj, 662, 389
		

%	
\bibitem[Hammonds et al.(2015)]{Hammonds15}
Hammonds, M., Mori, T., Usui, F., \& Onaka, T. 2015, \planss, 116, 73
%

	
\bibitem[Hehre et al.(1986)]{Hehre1986}
Hehre, W. J., Radom, L., Schleyer, P.V.R., et al., 1986 Ab initio molecular orbital theory (New York : Wiley) %ISBN 0471812412
	
\bibitem[Hertwig \& Koch(1997)]{hertwig1997} 
Hertwig, R. H., \& Koch, W. 1997 Chemical Physics Letters, 268, 345
	

	
\bibitem[Hrivnak et al.(2007)]{hrivnak07}
Hrivnak, B. J., Geballe, T. R., \& Kwok, S. 2007, \apj, 662, 1059
	

	
\bibitem[Hsia et al.(2017)]{hsia2017}
Hsia, C.-H., et al. 2017, in preparation

	

	
\bibitem[Jensen(2001)]{jensen2001}
Jensen, F. 2001, Journal of Chemical Physics, 115, 9113
	
\bibitem[Jensen(2002)]{jensen2002} 
Jensen, F. 2002, Journal of Chemical Physics, 116, 7372
	

	
\bibitem[Jourdain de Muizon et al.(1986)]{Jourdain86} 
Jourdain de Muizon, M., Geballe, T. R., D'Hendecourt, L. B., \& Baas, F. 1986, \apj, 306, L105
	
\bibitem[Jourdain de Muizon et al.(1990)]{Jourdain90}
Jourdain de Muizon, M., Cox, P., \& Lequeux, J. 1990, \aaps, 83, 337
	
\bibitem[Knacke(1977)]{knacke} 
Knacke, R. F. 1977, \nat, 269, 132

\bibitem[Kwok \& Zhang (2011)]{kwok2011}
Kwok, S., \& Zhang, Y. 2011, \nat, 479, 80

\bibitem[Kwok \& Zhang(2013)]{kwok2013}
Kwok, S., \& Zhang, Y. 2013, \apj, 771, 5



	
\bibitem[Laury et al.(2012)]{laury2012} 
Laury, M. L., Carlson, M. J., \& Wilson, A. K. 2012, Journal of Computational Chemistry, 33, 2380
	

	
\bibitem[Mattioda et al.(2017)]{Mattioda2017} 
Mattioda, A. L., Bauschlicher,  C. W., Ricca, A. et al. 2017 Spectrochimica Acta Part A: Molecular and Biomolecular Spectroscopy, 181, 286
	

\bibitem[Merrill et al.(1975)]{merrill1975} 
Merrill, K. M., Soifer, B. T., \& Russell, R. W. 1975, \apj, 200, L37
	
%\bibitem[Miani et al.(2000)]{Miani2000} 
%Miani, Andrea., Cane, Elisabetta., Palmieri, Paolo., et al. 2000, JCP, 112, 248
	

	
\bibitem[Minyaev \& Minkin(2008)]{Minyaev2008} 
Minyaev, R. M., \& Minkin, V. I. 2008 Russian Journal of General Chemistry, 78, 732 
		

	
%\bibitem[NASA-PAH(2015)]{NASA-PAH} 
%NASA-AMES PAH IR Spectral Database, http://www.astrochem.org/pahdb/
	
\bibitem[Oomens et al.(2006)]{Oomens2006}
Oomens, J.,  Polfer, N., Pirali, O., et al. 2006, Journal of Molecular Spectroscopy, 238, 158
	

	
\bibitem[Payne et al.(1997)]{Payne1977} 
Payne, P. W., \& Allen, L. C., 1977, in {\it Applications of Electronic Structure Theory (Modern Theoretical Chemistry series}, ed. Henry Schaefer (New York: Plenum), 29
	
	
\bibitem[Pirali et al.(2007)]{Pirali2007} 
Pirali, O., Vervloet, M. E. Dahl, J., et al. 2007, ApJ, 661, 919

\bibitem[Puetter et al.(1979)]{puetter1979}
Puetter, R. C., Russell, R. W., Willner, S. P., \& Soifer, B. T. 1979, \apj, 228, 118

\bibitem[Puget, \& Le{'}ger(1989)]{puget1989} 
Puget, J. L., \& Le{'}ger, A. 1989, \araa, 27, 161


\bibitem[Russell, Soifer, \& Willner(1977)]{RSW77} 
Russell, R. W., Soifer, B. T., \& Willner, S. P. 1977, \apj, 217, L149

\bibitem[Russell, Soifer, \& Merrill(1977)]{RSM77}
Russell, R. W., Soifer, B. T., \& Merrill, K. M. 1977, \apj, 213, 66	

\bibitem[Sadjadi et al.(2015a)]{SKZ2015-1} 
Sadjadi, S., Zhang, Y., \& Kwok, S. 2015a, \apj, 801, 34
	
\bibitem[Sadjadi et al.(2015b)]{SKZ2015-2}
Sadjadi, S., Zhang, Y., Kwok, S. 2015b, \apj, 807, 95

\bibitem[Sakata et al.(1987)]{sakata1987} 
Sakata, A., Wada, S., Onaka, T., \& Tokunaga, A. T. 1987, \apj, 320, L63	


\bibitem[Sandford et al.(1991)]{Sandford1991}
Sandford, S. A., Allamandola, L. J., Tielens, A. G., et al. 1991, \apj,  371, 607

\bibitem[Scott, \& Duley(1996)]{scott1996} 
Scott, A., \& Duley, W. W. 1996, \apj, 472, L123	

\bibitem[Schutte et al.(1993)]{sch93} 
Schutte, W.~A., Tielens, A.~G.~G.~M., \& Allamandola, L.~J.\ 1993, \apj, 415, 397

	
\bibitem[Socrates(2001)]{Socrates2001}
Socrates, G.  2001 {\em Infrared and raman characteristic group frequencies Tables and Charts} (Chichester: Wiley)
	
\bibitem[Song et al.(2003)]{Song03}
Song, I.-O., Kerr, T. H., McCombie, J. \& Sarre, P. J. 2003, \mnras, 346, L1
	


\bibitem[Tokunaga et al.(1991)]{tokunaga1991}
Tokunaga, A. T., Sellgren, K., Smith, R. G., et al. 1991, \apj, 380, 452
	
\bibitem[van Diedenhoven et al.(2004)]{van04} 
van Diedenhoven, B., Peeters, E., Van Kerckhoven, C., et al.\ 2004, \apj, 611, 928

\bibitem[Wada \& Tokunaga(2006)]{Wada2006} 
Wada, Setsuko., \& Tokunaga, Alan T.  2006, Natural Fullerences and Related Structures of Elemental Carbon, Chapter 3, 31
	
\bibitem[Wagner et al.(2000)]{Wagner00}
Wagner, D. R., Kim, H. S., \& Saykally, R. J. 2000, \apj, 545, 854
	



\bibitem[Woolf, \& Ney(1969)]{woolf1969} 
Woolf, N. J., \& Ney, E. P. 1969, \apj, 155, L181

\bibitem[Yang et.al(2013)]{Yang2013}
Yang, X. J., Glaser, R., Li, A., \& Zhong, J. X. 2013, \apj, 776, 110 

\bibitem[Zhang \& Kwok(2013)]{zhang2013}
Zhang, Y., \& Kwok, S. 2013, Earth, Planets, and Space, 65, 1069
	
\bibitem[Zvereva et al.(2011)]{Zvereva2011}
Zvereva, E. E., Shagidullin, A. R., Katsyuba, S. A. 2011, Journal of Physical Chemistry, 115, 63
	
	
\end{thebibliography}
\end{document}